\documentclass[prb,floatfix,twocolumn,amsmath,amssymb,amsfonts,superscriptaddress]{revtex4}
\usepackage{graphicx,subfigure}
\usepackage[usenames,dvipsnames]{color}
\usepackage[dvips]{epsfig}

\usepackage{verbatim}

\begin{document}

\title{Magnetic domain wall motion in a nanowire: depinning and creep}
\author{Jisu Ryu}
\affiliation{PCTP and Department of Physics, Pohang University of
Science and Technology (POSTECH), Pohang, Kyungbuk 790-784,
Republic of Korea}
\author{Sug-Bong Choe}
\email{sugbong@snu.ac.kr} \affiliation{Center for Subwavelength
Optics and Department of Physics, Seoul National University, Seoul
151-742, Republic of Korea}
\author{Hyun-Woo Lee}
\email{hwl@postech.ac.kr} \affiliation{PCTP and Department of
Physics, Pohang University of Science and Technology (POSTECH),
Pohang, Kyungbuk 790-784, Republic of Korea}

\begin{abstract}
The domain wall motion in a magnetic nanowire is examined
theoretically in the regime where the domain wall driving force is
weak and its competition against disorders is assisted by thermal
agitations. Two types of driving forces are considered; magnetic
field and current. While the field induces the domain wall motion
through the Zeeman energy, the current induces the domain wall
motion by generating the spin transfer torque, of which effects in
this regime remain controversial. The spin transfer torque has two
mutually orthogonal vector components, the adiabatic spin transfer
torque and the nonadiabatic spin transfer torque. We investigate
separate effects of the two components on the domain wall
depinning rate in one-dimensional systems and on the domain wall
creep velocity in two-dimensional systems, both below the Walker
breakdown threshold. In addition to the leading order contribution
coming from the field and/or the nonadiabatic spin transfer
torque, we find that the adiabatic spin transfer torque generates
corrections, which can be of relevance for an unambiguous analysis
of experimental results. For instance, it is demonstrated that the
neglect of the corrections in experimental analysis may lead to
incorrect evaluation of the nonadiabaticity parameter. Effects of
the Rashba spin-orbit coupling on the domain wall motion are also
analyzed.
\end{abstract}

\maketitle

\section{introduction}
%DW creep motion, arrehenius law
A magnetic domain wall (DW) in a ferromagnetic nanowire is an
important subject in spintronics. A new type of logic device is
proposed~\cite{allw05} based on the DW dynamics and a DW-based memory
is also proposed~\cite{park08}, which may have merits such as
nonvolatility, high speed, high density, and low power
consumption.

The dynamics of a DW varies considerably depending on the relative
strength of DW driving forces (such as a magnetic field and a
current) with respect to disorders, which tend to suppress the DW
motion. If the forces are sufficiently strong or the disorders
are sufficiently weak~\cite{Ryu2009JAP}, the DW dynamics does not
deviate much from the ideal dynamics in the absence of disorders.
While some experiments~\cite{yamg04, ver04,yamn04,kla05} are estimated to be in this
regime, many other experiments~\cite{yamn07, moor08,leme98,kjkim09} appear to be in the
regime where the disorders are important. It is thus desired to
understand the DW dynamics in the weak driving force regime where
the competition between the DW driving forces and the disorders is
significant.

The DW motion in the weak driving force regime is an important
example in the field of driven interfaces. The study on driven
interfaces has a long history~\cite{Barabasi1995Book} and
addresses many physical systems such as surface growth of a
crystal~\cite{Jullien1992Book}, vortex line motion in high
temperature superconductors~\cite{Blatter1994RMP}, and fluid
propagation in porous media~\cite{Rubio1989PRL}. Through a long
series of theoretical works~\cite{natt90,Blatter1994RMP,chau00}, a
simple picture has emerged; the interface motion becomes
collective and the collective length scale $L_{\rm col}$~\cite{kjkim09}, which
characterizes the length scale of collectively moving interface
segments, diverges in the weak driving force limit. Due to the
divergence of $L_{\rm col}$, the interface has to overcome an
increasingly larger energy barrier as the driving force becomes
weaker, with the energy barrier $E_B$ as a function of the driving
force $f$ diverging as a power law, $E_B \propto f^{-\mu}$
($\mu>0$). Interestingly the creep exponent $\mu$ is universal in
the sense that its value does not change continuously with
variations of system details and is affected only by a small
number of key features such as the system dimensionality.
Systems with the same exponent are said to be in the same
universality class.

This prediction has been unambiguously confirmed for
the field-driven DW motion in metallic ferromagnets~\cite{leme98}, where the DW
velocity $v$ is proportional to $\exp(-\kappa H^{-\mu}/k_B T)$.
Here $k_B$ is the Boltzmann constant, $T$ is the temperature, $H$
is the magnetic field strength, and $\kappa$ is a constant. Note that this
behavior of $v$ is a combined result of the power law scaling of
the energy barrier $E_B=\kappa H^{-\mu}$ and the Arrhenius law~\cite{Blatter1994RMP} $v\propto\exp(-E_B/k_B T)$. The creep exponent $\mu$ is found
to be $\approx 0.25$, which agrees with the theoretically
predicted value $1/4$ in two-dimensional (2D)
systems~\cite{chau00}.

A pioneering experiment~\cite{yamn07} revealed
interesting twists. For nanowires made of a ferromagnetic
semiconductor (Ga,Mn)As, the energy barrier for the field-driven
DW motion was found to scale as $H^{-\mu}$, where $\mu\approx 1.2$
instead of $1/4$. This difference was attributed to the different
nature of disorders; while disorder potential energy is
short-range correlated in metallic ferromagnets, it was argued
that in ferromagnetic semiconductors, disorder {\it force} is
short-range correlated. Since the disorder potential energy is
obtained by integrating the disorder force, it implies that the
disorder potential energy is then long-range
correlated~\cite{chau00}. For such cases, it is
known~\cite{chau00} that the nature of the correlation along the
DW segments is modified and the value of $\mu$ indeed changes.

Another interesting twist of the
experiment~\cite{yamn07} is that for the
current-driven DW motion, the effective energy barrier was
reported to scale as $J^{-\mu}$, where $J$ is the current density
and $\mu\approx 0.33$ rather than $1/4$ or $1.2$. Thus two
different creep exponent values ($1.2$ and $0.33$) were obtained from
the {\it same} material, implying that the current-driven DW
motion is qualitatively different from the field-driven DW motion.

It is believed that the current induces the DW motion in a
nanowire by generating the spin transfer torque (STT). The STT has two
mutually orthogonal vector components, the adiabatic STT and the
nonadiabatic STT~\cite{zhan04,thia05}. The nonadibatic STT has
 similar properties as the magnetic field while the adiabatic STT has
very different properties. Thus the experimental
result~\cite{yamn07} implies that the nonadiabatic
STT cannot be the main driving force of the DW motion. In fact the
exponent $\mu\approx 0.33$ has been
interpreted~\cite{yamn07} as an indication that the
current-driven DW motion is mainly due to the adiabatic STT.

This interpretation is at odds, though not contradictory, with
other results. In metallic ferromagnets, the onset of the
adiabatic-STT-driven DW motion is estimated~\cite{tata04, zli04} to occur
at the current density of $\sim10^{9}$ A/cm$^2$, which is
unendurably high for most experimental systems. Thus the DW motion
realized at lower current densities are usually attributed to the
nonadibatic STT.

This situation strongly motivates
experimental~\cite{Lee01preprint} and theoretical~\cite{jvkim09, duin08, luca09}
studies of the current-driven DW motion in metallic ferromagnets.
This paper aims at theoretical explorations of this issue based on
the observation that the DW anisotropy, characterizing the energy
cost associated with the change in the tilting angle of the magnetization inside a DW
, is orders of magnitude larger in metallic ferromagnets than in ferromagnetic semiconductors. Since the DW anisotropy tends to suppress variations
of the tilting angle, we assume that the DW creep motion in
metallic ferromagnets exhibits the below-the-Walker-breakdown-like
behavior in the sense that the amplitude of the tilting angle
variations during the creep motion stays much smaller than $2\pi$.
For the field-driven DW creep motion, this assumption is
experimentally supported since the experimental value $\sim 0.25$
of the creep exponent agrees with the prediction $1/4$ of the
theory~\cite{chau00}, in which the tilting angle dynamics is
completely suppressed. For the current-driven DW creep motion, the
assumption requires an experimental confirmation. A recent
experiment~\cite{Lee01preprint} reports the purely current-driven
DW creep motion in metallic ferromagnets. For ferromagnetic semiconductors, in contrast, it appears that the assumption may not be valid. For the
current-driven DW creep motion, it was
argued~\cite{yamn07} that each thermally-assisted
tunneling event overcomes the energy barrier generated by the DW
anisotropy, implying that each tunneling event is accompanied by
the tilting angle change by $\sim \pi$.

The paper is structured as follows. In Sec.~\ref{sec:1D}, we
discuss the DW depinning from a single potential well in
one-dimensional (1D) systems. Analysis of this relatively simple
problem clearly illustrates separate roles of the magnetic field,
the adiabatic STT, and the nonadiabatic STT on the
thermally-assisted tunneling of a DW. It also allows one to
identify relevant factors affecting the tunneling, which therefore
should be included in the analysis. In this sense, Sec.~\ref{sec:1D} is
pedagogical. Nevertheless predictions in Sec.~\ref{sec:1D} can be tested
in real experiments since a DW exhibits the 1D dynamics~\cite{kjkim09} when
$L_{\rm col}$ becomes larger than both the thickness and width of
a nanowire. In particular, it is predicted that when the depinning
rate is used as a tool to evaluate the nonadiabaticity
parameter~\cite{zhan04,thia05}, characterizing the strength of the
nonadibatic STT, it may lead to incorrect values if disorders in a
nanowire have certain features. In Sec.~\ref{sec:2D}, the DW creep
motion in 2D systems is analyzed. Separate roles of the magnetic
field, the adiabatic STT, and the nonadiabatic STT on the creep
motion are clarified. In addition to the leading order
contribution to the creep motion in the vanishing DW driving force limit,
next leading order contributions are also obtained. Although the
next leading order contributions are irrelevant as far as the
theoretical determination of the creep exponent and the
universality class is concerned, they may nevertheless be relevant
in {\it experimental} determination of the creep exponent since
experiments are always performed at small but {\it finite} driving
force strength. At the end of both Secs.~\ref{sec:1D} and
\ref{sec:2D}, effects of the Rashba spin-orbit coupling (RSOC) are
discussed. The emergence of the RSOC in ferromagnetic nanowires is
recently demonstrated~\cite{Miron2010NatMat}.
Section~\ref{sec:conclusion} concludes this paper.

%%%%%%%%%%%%%%%%%%%%%%%%%%%%%%%%%%%%%%%%%%
\section{DW depinning in 1D}\label{sec:1D}
%%%%%%%%%%%%%%%%%%%%%%%%%%%%%%%%%%%%%%%%%%
When both the thickness and the width of a magnetic nanowire are
sufficiently smaller than the collective length $L_{col}$, the
system reduces to a 1D problem and the configuration of a DW can
be described by two variables, the DW position $q$ and the
tilting angle $\psi$. This Section examines the DW depinning from
a potential well in this 1D regime.

%%%%%%%%%%%%%%%%%%%%%%%%%%%%%%%%%%%%%%%%%%%%%
\subsection{Effective energy}\label{sec:1DEb}
%%%%%%%%%%%%%%%%%%%%%%%%%%%%%%%%%%%%%%%%%%%%%
%%%%%%%%%%%%%%
\begin{figure}
\subfigure[]
{\includegraphics[width=0.2\textwidth]{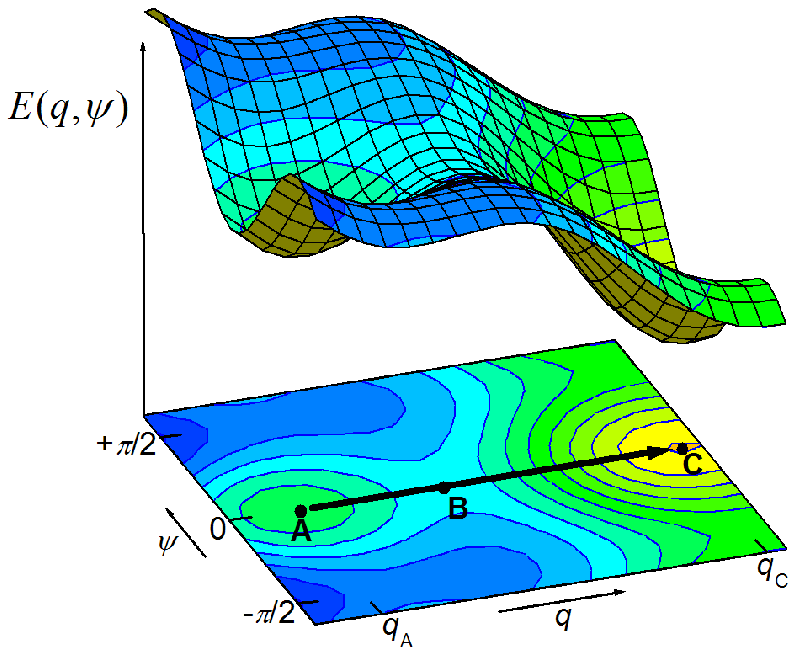}}
\subfigure[]
{\includegraphics[width=0.2\textwidth]{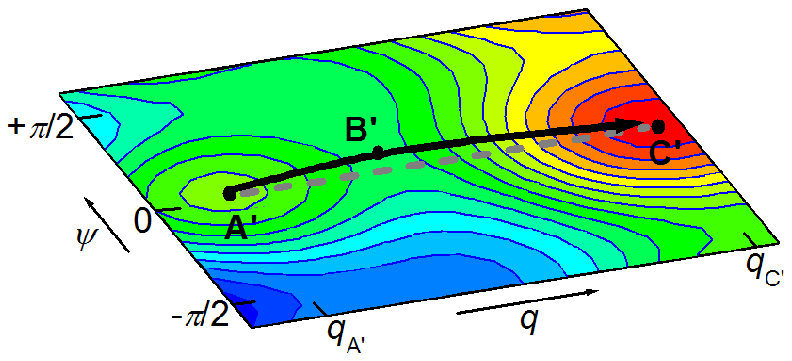}}
\caption{(Color online) Schematic plot of the energy
landscape $E(q,\psi)$ as a function of $q$ and $\psi$. (a) The
activation path (black solid arrow) of the thermally-assisted
transition for $H\neq0$ and $J=0$ is shown schematically from one local
minimum {\bf A} $[(q_A,\psi_A)=(q_{G0},\psi_{G0})]$ to another
local minimum {\bf C} $(q_C,\psi_C)$ through the saddle point {\bf
B} $[(q_B,\psi_B)=(q_{S0},\psi_{S0})]$. For simplicity,
$\psi_{G0}=\psi_{S0}=\psi_C=0$ is assumed, and
$E(q_{G0},\psi_{G0})>E(q_{C},\psi_{C})$ is also assumed in this
plot. (b) When $J$ is turned on, the activation path
(black solid arrow) is deformed to {\bf A'}-{\bf B'}-{\bf C'}.
Here the coordinates of {\bf A'} and {\bf B'} correspond to
$(q_G,\psi_G)$ and $(q_S,\psi_S)$ in the text, respectively. Note
that the path is now curved due to the adiabatic STT when
$\nu_{G}^{-2}\neq \nu_S^{-2}$.}\label{fig:1Dmodel}
\end{figure}
%%%%%%%%%%%%
%1D motion description
In the 1D regime, the response of the DW collective coordinates
$(q,\psi)$ to an external magnetic field $H$ and/or an electric
current of density $J$ is described by the following equations,
\begin{align}
\alpha\frac{\dot q}{\lambda}-\dot\psi &= \gamma_0(H-\beta\chi J)
- \frac{\gamma_0}{2M_S\Omega}\frac{\partial V}{\partial q},\label{eq:1DEOM1}\\
\frac{\dot q}{\lambda}+\alpha\dot\psi &= -\gamma_0\chi
J-\frac{\gamma_0}{2M_S\Omega\lambda}\frac{\partial
V}{\partial\psi}\label{eq:1DEOM2},
\end{align}
where $\alpha$ is the Gilbert damping parameter, $\lambda$ is the
DW width, $\gamma_0$ is the gyromagnetic ratio, $M_S$ is the
saturation magnetization, $\Omega$ is the cross-sectional area of
a nanowire, $V(q,\psi)$ is the DW potential energy, and the
dimensionless parameter $\beta$ is the nonadiabaticity
coefficient~\cite{zhan04,thia05} representing the strength of the
nonadiabatic STT. $\chi=\hbar P/2\lambda eM_S$ $(<0)$ % Since \dot q>0 for J>0 in this convention, \chi<0.
is a constant with the dimension $H/J$, $P$ is the
spin-polarization of the current, and $\hbar$ is the Planck
constant. In Eqs.~(\ref{eq:1DEOM1}) and (\ref{eq:1DEOM2}), the
sign convention of $J$ is chosen in such a way that positive $J$
drives the DW towards the positive $q$ direction. On the other
hand, the sign convention of $H$ should depend on the types of the
DW; In nanowires with the in-plane anisotropy, for instance,
opposite signs should be adopted for the head-to-head and
tail-to-tail DWs. Below for simplicity, we consider one particular
sign only, and assume that the positive $H$ tends to drive the DW
towards the positive $q$ direction.

The DW potential energy $V(q,\psi)$ consists of the DW anisotropy
energy $2\Omega \lambda K_d \sin^2 \psi$ and a disorder potential
energy, where $K_d$ represents the strength of the DW anisotropy.
Here $\psi$ is defined in a way that $\psi=0$ for the tilting
angle preferred by the DW anisotropy. When the disorder potential
energy depends only on $q$, Eqs.~(\ref{eq:1DEOM1}) and
(\ref{eq:1DEOM2}) become equivalent to Eqs.~(3) and (4) in
Ref.~\onlinecite{swju08}.  In general, however, the disorder potential
energy may also depend on $\psi$.

%{\bf [This paragraph has been revised. So please check carefully]}
%
Such $\psi$ dependence of $V$ may arise in various ways. For
instance, the value of $M_S$ may fluctuate from position to
position. Recalling that $K_d$ depends~\cite{swju08} on $M_S$,
$K_d$ can then be decomposed into its spatial average part and the
fluctuating part $\delta K_d(q)$. The fluctuating part of the DW
anisotropy energy [$\propto \delta K_d(q) \sin^2 \psi$] may be
absorbed to $V$ to generate its $\psi$ dependence. Similar
dependence may arise from the position-to-position fluctuation of
$\Omega$. In these types of disorder, the preferred tilting angle
remains unaffected and only the strength of the DW anisotropy
fluctuates.
Some types of disorder may generate opposite effects. In a
magnetic nanowire with the perpendicular magnetic anisotropy
(PMA), the interface between the magnetic layer and the neighboring
layer plays important roles for the anisotropy. When the interface is not perfectly flat
and becomes rough\cite{mli98, CHchang94, zhao99,vaz09}, the
preferred anisotropy direction fluctuates from position to
position. In this case, the preferred tilting angle fluctuates
while the strength of the DW anisotropy may not fluctuate.

Below we consider this general situation, in which the disorder
potential energy depends both on $q$ and $\psi$. In
Ref.~\onlinecite{jvkim09}, the $\psi$ dependence of the disorder
potential energy is included in its initial formulation but
ignored when the depinning rate is calculated. We demonstrate
below that the $\psi$ dependence of $V$ generates interesting
consequences.

Based on the Lagrangian formulation, Eqs.~(\ref{eq:1DEOM1}) and
(\ref{eq:1DEOM2}) may be considered as the Lagrange's equations of
the Lagrangian $L$ and the dissipation function $F$,
\begin{align}
L= & \frac{M_S \Omega}{\gamma_0}(q\dot\psi-\dot{q}\psi)-V(q,\psi)
\label{eq:Lagrangian}
\\
& +2M_S \Omega q (H-\beta \chi J)-2M_S \Omega \lambda \psi \chi J,
\nonumber \\
F = & \alpha \frac{M_S \Omega}{\gamma_0 \lambda}(\dot{q}^2+\lambda^2
\dot{\psi}^2). \label{eq:Dissipation}
\end{align}
The Lagrangian in Eq.~(\ref{eq:Lagrangian}) is then transformed to
the Hamiltonian {\it i.e.} the effective energy function $E$,
\begin{align}\label{eq:1DEn}
E(q,\psi) =& V(q,\psi)-2M_S\Omega q(H-\beta\chi J) \\
& +2M_S\Omega\lambda\psi\chi J. \nonumber
\end{align}
Here we have used the term {\it effective} energy since $E$ is
{\it not} a single valued function\cite{stil07} in the sense that
$E(q,\psi)\neq E(q,\psi+2\pi)$ although $(q,\psi)$ and
$(q,\psi+2\pi)$ represent the same magnetic configuration. Thus
some care should be exercised when Eq.~(\ref{eq:1DEn}) is used to
analyze the DW dynamics above the Walker breakdown threshold,
where $\psi$ changes more than $2\pi$. Below the Walker breakdown
threshold, on the other hand, the dynamics of $\psi$ is confined
to a value range narrower than $2\pi$ and $E(q,\psi)$ can be
regarded as a single valued function.

%%%%%%%%%%%%%%%%%%%%%%%%%%%%%%%%%%%%%%%%%%%%%%%%%%%%%%%%%%%%%%%%%%%%%%
\subsection{Effective energy barrier} \label{sec:1D Energy barrier}
%%%%%%%%%%%%%%%%%%%%%%%%%%%%%%%%%%%%%%%%%%%%%%%%%%%%%%%%%%%%%%%%%%%%%%
Figure~\ref{fig:1Dmodel} shows schematically the energy profile,
to which the DW is subject. The DW has to overcome an energy barrier to get depinned from a given
potential well. When the DW driving force ($H$ or $J$) is small, the height of the
energy barrier is sufficiently higher than the DW energy measured
from the bottom of the potential well and the DW overcomes the
large energy barrier by exploiting the thermal agitation.  Thus
the depinning time from potential wells is governed (within the
exponential accuracy) by the energy barrier via the Arrhenius law.
When the depinning time is much longer than the relaxation time
inside potential wells, the energy barrier is defined as the
difference between the saddle point energy and the local ground
state energy.
One remark is in order.
%Secondly, we remark that the thermal
%fluctuation effects are ignored in Eqs.~(\ref{eq:1DEOM1}) and
%(\ref{eq:1DEOM2}) despite their importance in the DW depinning. We
%intentionally neglected them in the derivation of the energy
%function $E$ since the thermal fluctuation effects can be
%accounted for by using the Arrhenius law {\bf [Reference]}
%provided that the thermal fluctuations satisfy the
%fluctuation-dissipation theorem {\bf [Reference needed]}.
While the Arrhenius law is based on the fluctuation-dissipation
theorem\cite{brown63, kubo70, foros08},
the theorem does not generally hold when $J$ is finite and the system is
thus in nonequilibrium situations. However it has been demonstrated that for small
$J$~\cite{Duine2007PRB} and below the Walker breakdown regime~\cite{Kim2010PRB},
 thermal fluctuations still satisfy the theorem, justifying the use of
the Arrhenius law in this case.

%saddle and ground shifts, E_B
The energy barrier $E_B$ depends on $H$ and $J$, and we examine
this dependence. For $H=J=0$, $E(q,\psi)$ reduces to $V(q,\psi)$.
Let $(q_{G0}, \psi_{G0})$ and $(q_{S0}, \psi_{S0})$ denote
respectively the local ground state and saddle point
configurations of $V(q,\psi)$. Note that we  introduce separate
parameters $\psi_{G0}$ and $\psi_{S0}$. Although $\psi_{S0}-\psi_{G0}$ will be much smaller than $2\pi$ in the regime below the Walker breakdown, the difference is nonzero
in general due to the $\psi$ dependence of the disorder potential energy.
To examine effects of small $H$ and $J$, $V(q,\psi)$ may be Taylor
expanded near these configurations;
\begin{equation} \label{eq:Taylor ground}
V \approx \omega_G^2(q-q_{G0})^2+\nu_G^2(\psi-\psi_{G0})^2,
\end{equation}
for $(q,\psi)$ near $(q_{G0}, \psi_{G0})$, and
\begin{equation} \label{eq:Taylor saddle}
V \approx V_0-\omega_S^2(q-q_{S0})^2+\nu_S^2(\psi-\psi_{S0})^2,
\end{equation}
for $(q,\psi)$ near $(q_{S0}, \psi_{S0})$. Here $\omega_{G/S}$ and
$\nu_{G/S}$ are the potential stiffness, and the potential depth
$V_0$ amounts to the energy barrier height for $H=J=0$. Note that
$\omega_G^2$ in Eq.~(\ref{eq:Taylor ground}) and $\omega_S^2$ in
Eq.~(\ref{eq:Taylor saddle}) appear with the opposite signs due to
the difference between the local ground state and saddle point
(Fig.~\ref{fig:1Dmodel}). Note also that we distinguish $\nu_G$
and $\nu_S$ in order to take account of the $\psi$ dependence of
the disorder potential energy.

The driving forces $H$ and $J$ modify the local ground state and
saddle point configurations to, say, $(q_G,\psi_G)$ and
$(q_S,\psi_S)$. For small $H$ and $J$, the modified configurations
can be determined from $\delta E=0$ with the aid of
Eqs.~(\ref{eq:Taylor ground}) and (\ref{eq:Taylor saddle}). One
obtains
\begin{align}
q_G &= q_{G0}+{M_S\Omega}{\omega_{G}^{-2}}(H-\beta\chi J), \label{eq:1DdqG} \\
\psi_{G} & =\psi_{G0}-{M_S\Omega}{\nu_G^{-2}}\lambda\chi J, \label{eq:1DdpsiG} \\
q_{S}&=q_{S0}-{M_S\Omega}{\omega_{S}^{-2}}(H-\beta\chi J), \label{eq:1DdqS}\\
\psi_{S}&=\psi_{S0}-{M_S\Omega}{\nu_S^{-2}}\lambda\chi J,
\label{eq:1DdpsiS}
\end{align}
The evaluation of the energy barrier
$E_B=E(q_S,\psi_S)-E(q_G,\psi_G)$ is now trivial. One finds,
\begin{align}\label{eq:1DEb}
E_B - V_{0}= &   -2 \delta q_0 M_S\Omega (H-\beta\chi J) \\
& +2\delta \psi_0 M_S\Omega(\lambda\chi J) \nonumber\\
& +{M_S^2\Omega^2}{\omega_+^{-2}}(H-\beta\chi J)^2
\nonumber \\
& -{M_S^2\Omega^2}{\nu_-^{-2}}(\lambda\chi J)^2, \nonumber
\end{align}
where $\delta q_0 \equiv q_{S0}-q_{G0}$, $\delta \psi_0\equiv
\psi_{S0}-\psi_{G0}$, $\omega_+^{-2}\equiv
\omega_S^{-2}+\omega_G^{-2}$ and $\nu_-^{-2}\equiv
\nu_S^{-2}-\nu_G^{-2}$. Equation~(\ref{eq:1DEb}) clearly shows the
effect of $H$ and $J$ on the energy barrier.
Among the two components of the STT produced by $J$, the
non-adiabatic STT ($\propto \beta\chi J$) in the first and third
lines of Eq.~(\ref{eq:1DEb}) has the exactly same effect as the
magnetic field $H$ while the second and the fourth lines of
Eq.~(\ref{eq:1DEb}) indicate that the effect of the adiabatic STT
($\propto \lambda\chi J$) is qualitatively different from the
field effect.
%For a magnetic system with a strong uniaxial magnetic
%anisotropy, a DW polarity reversal during the creep motion is
%hardly expected. Thus $\psi_{S0}-\psi_{G0}=0$ during the motion.
%Then, the $O(J)$ contribution to $E_B$ arises solely from the
%non-adiabatic STT.

The depinning rate $1/\tau$ from a potential well is then given by
\begin{equation} \label{depinning rate}
\frac{1}{\tau}=\frac{1}{\tau_0}\exp\left[-\frac{E_B(H,J)}{k_BT}\right],
\end{equation}
where $1/\tau_0$ amounts to the attempt frequency and the $H$
and $J$ dependence of $E_B$ are given in Eq.~(\ref{eq:1DEb}).

%comparison with JVkim
Recently Kim and Burrowes~\cite{jvkim09} analyzed the effective
energy barrier for the purely current-driven DW creep motion in
1D. Equation~(39) in their work indicates that $J$ modifies the
energy barrier $E_B$ through a linear term ($\propto \beta J$),
 and a quadratic term ($\propto
\beta^2 J^2$), both of which arise from the nonadiabatic STT. Our
result [first and third lines in Eq.~(\ref{eq:1DEb})] agrees with
this result as far as these two terms are concerned. However our
result predicts that there are another linear term [$\propto
\lambda\chi J$, the second line in Eq.~(\ref{eq:1DEb})] and 
quadratic term [$\propto \lambda^2 \chi^2 J^2$, the fourth line in
Eq.~(\ref{eq:1DEb})], which arise from the adiabatic STT.
% with the $\psi$
%dependence of the disorder potential energy.
%The last term in Eq.~(\ref{eq:1DEb}) amounts to this
%additional quadratic term.
This difference between our result and Ref.~\onlinecite{jvkim09} stems
from the nature of the disorders; In Ref.~\onlinecite{jvkim09}, the
calculation of $E_B$ assumed that the disorder contribution to
$V(q,\psi)$ depends only on $q$ and does not depend on $\psi$,
whereas we consider more realistic situations where the disorder
contribution depends not only on $q$ but also on $\psi$. This
dependence on $\psi$ appears in the second and last terms in
Eq.~(\ref{eq:1DEb}) through the factors $\delta \psi_0$ and
$\nu_-^{-2}$.

DW depinning experiments\cite{burr09, alva10, elts10, boul08,
hein10} are sometimes used as a tool to determine the
nonadiabaticity parameter $\beta$. When the $\psi$ dependence of
the disorder potential energy is negligible and thus $\delta
\psi_0=\nu_-^{-2}=0$, one can verify from Eq.~(\ref{eq:1DEb}) that
$E_B$ depends on $H$ and $J$ through a single variable $H-\beta
\chi J$. Thus by comparing the ``efficiency" of $H$ and $J$ in the
DW depinning, one can determine $\beta$. In general, however, the
$\psi$ dependence of the disorder potential energy may not be
negligible. In such situations and in the limit $H,J\rightarrow
0$, the $H$ and $J$ dependence of $E_B$ appears through a
different single variable $H-\beta \chi J - \lambda\chi J \delta
\psi_0/\delta q_0$ thus uncareful experimental evaluation may
incorrectly identify
\begin{equation} \label{eq:beta'}
\beta'=\beta+\lambda\frac{\delta \psi_0}{\delta q_0}
\end{equation}
as $\beta$. Thus possible $\psi$ dependence of the disorder
potential energy should be carefully examined for the correct
evaluation of $\beta$.

As discussed in Sec.~\ref{sec:1DEb}, the $\psi$ dependence of $V$
may be qualitatively different depending on details of disorders.
When $\nu_-^{-2}\neq 0$ but $\delta \psi_0= 0$, the second
contribution in Eq.~(\ref{eq:beta'}) vanishes, simplifying the
experimental evaluation of $\beta$.
When $\nu_-^{-2}= 0$ but $\delta \psi_0 \neq 0$, on the other
hand, the second contribution in Eq.~(\ref{eq:beta'}) may not be
negligible.
%We consider two experimental situations. Firstly, we consider the
%situation when position-by-position fluctuations of the magnetic
%anisotropy strength is the origin of the $\psi$ dependence of the
%disorder potential energy. Then $\nu_-^{-2}\neq 0$ but $\delta
%\psi_0= 0$. In this case, the second contribution in
%Eq.~(\ref{eq:beta'}) may be ignored, simplifying the experimental
%evaluation of $\beta$.
%%
%Secondly, we consider the situation when the preferred direction
%of the magnetization fluctuates position-by-position and this is
%the source of the $\psi$ dependence of the disorder potential
%energy. Then, $\nu_-^{-2}=0$ but $\delta \psi_0 \neq 0$. In this
%case, the second contribution in Eq.~(\ref{eq:beta'}) may not be
%negligible.
%
A possible way to avoid the incorrect evaluation of $\beta$ in
this case is to take an average of $\beta'$ for multiple potential
wells. Since the sign of $\delta \psi_0$ is expected to fluctuate
from potential wells to potential wells, this averaging process
may be able to remove the second contribution of $\beta'$
proportional to $\delta \psi_0$.
%(see the Sec.~\ref{sec:1D creep
%velocity} for a related discussion) {\bf [Remove the referral to
%Sec.~\ref{sec:1D creep velocity} for the arXiv version]}.
By the way, the sign fluctuations of $\delta q_0$ can be
suppressed in this averaging process since the depinning to the
right ($\delta q_0>0$) and to the left ($\delta q_0<0$) are distinguishable in experiments.

Lastly we compare two contributions [third and fourth lines in
Eq.~(\ref{eq:1DEb})], both of which generate the $J$-quadratic
contributions to $E_B$. They have one important difference; the
third line, which arises from the nonadiabatic STT, always enhance
$E_B$ and thereby lower the depinning rate while the fourth line,
which arises from the adiabatic STT, may either increase or
decrease $E_B$ since $\nu_-^{-2}$ can be positive or negative
depending on the nature of disorders.  Thus in case that experiments
find the $J$-quadratic contribution enhances the depinning rate,
it implies that the adiabatic STT makes a larger contribution to
the $J$-quadratic dependence of $E_B$ than the nonadiabatic STT.

%we remark that $\nu_-^{-2}$ can have in principle both signs. This
%results in a qualitative difference between the two quadratic
%contributions [third and fourth lines in Eq.~(\ref{eq:1DEb})] in
%$J$; In case of a purely current-driven DW depinning, the third
%line in Eq.~(\ref{eq:1DEb}), which arises from the nonadiabatic
%STT, always enhance $E_B$ and thereby lower the depinning rate
%while the fourth line, which arises from the adiabatic STT, may
%lower $E_B$. Thus in case experiments find the quadratic
%contribution in $J$ to enhance the depinning rate, it implies that
%the adiabatic STT makes a larger contribution to the quadratic
%$J$-dependence of $E_B$ than the nonadiabatic STT.

%A natural question is whether it is possible to distinguish in
%experiments the two types of quadratic terms in $J$. This may not
%be easy since experimental determination of $\omega_S$,
%$\omega_G$, $\nu_S$, and $\nu_G$ may be nontrivial. Below we
%demonstrate that a clear distinction is possible through the
%concept of the effective magnetic field.

%%%%%%%%%%%%%%%%%%%%%%%%%%%%%%%%%%%%%%%%%%%%%%%%%%%%%
\subsection{Effective magnetic field}\label{sec:1DH*}
%%%%%%%%%%%%%%%%%%%%%%%%%%%%%%%%%%%%%%%%%%%%%%%%%%%%%
%%%%%%%%%%%%%%
\begin{figure}
\subfigure[]
{\includegraphics[width=0.2\textwidth]{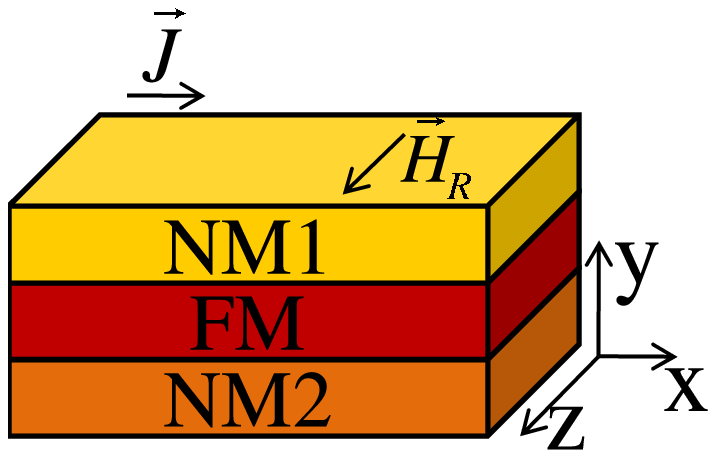}}
\subfigure[]
{\includegraphics[width=0.2\textwidth]{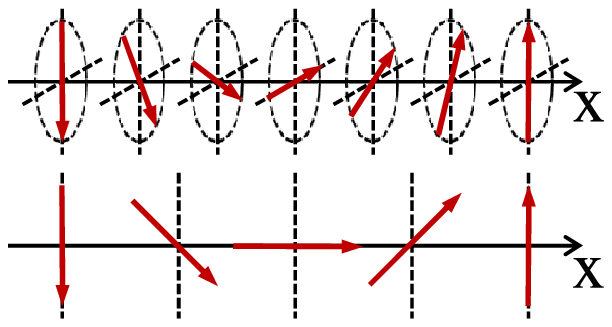}}
\caption{(Color online) (a) An example with the broken inversion
symmetry. The ferromagnetic layer (FM) is sandwiched between two
different nonmagnetic layers (NM1 and NM2), so that the inversion
symmetry is broken along the $\hat y$ direction. When the current
is injected along $\hat x$ direction, the Rashba spin-orbit
coupling makes the magnetization feel as if a magnetic field $\vec
H_R$ is applied~\cite{manc08} along the $\hat z$ direction. (b)
Schematic plots of the magnetization configurations for the Bloch
wall (upper plot) and the N\'eel wall(lower) in nanowires with the
perpendicular magnetic anisotropy. Solid arrows (colored in red)
represent local magnetic moments inside of a DW. For a Bloch
(N\'eel) wall, $\psi$ represents the angle between the magnetization and the positive $\hat z$($\hat x$) axis within the $xz$ plane.}\label{fig:rsocfield}
\end{figure}
%%%%%%%%%%%%
The DW depinning for the purely field-driven case is relatively
well understood~\cite{leme98,chau00}. Thus if one can ``map''
general situations with both $H$ and $J$ to the purely
field-driven case, it may provide a useful way to describe
experimental results in general situations.
%The effective magnetic field is a useful tool to summarize the
%experimental results in general situations with both $H$ and $J$
%nonzero.
The effective magnetic field is  one way to make this connection.
We define the effective field $H^*(H,J)$ of the DW depinning by
the relation $E_B(H^*,0)=E_B(H,J)$. $H^*(H,J)$ can be
experimentally extracted, for instance, from contour plots~\cite{Lee01preprint} of the
DW depinning rate as a function of $H$ and $J$.
From Eq.~(\ref{eq:1DEb}), one finds that $H^*$ satisfies
\begin{align}\label{eq:1DH1}
& H^{*2}-\frac{2\omega_+^{2}\delta q_0}{M_S\Omega}H^* \\
= & (H-\beta\chi J)^2-\frac{2\omega_+^{2}\delta q_0}{M_S\Omega}(H-\beta\chi J)\nonumber \\
& +\frac{2\omega_+^{2}\delta \psi_0}{M_S\Omega}(\lambda\chi
J)-\frac{\omega_+^{2}}{\nu_-^{2}}(\lambda\chi J)^2. \nonumber
\end{align}
Solving Eq.~(\ref{eq:1DH1}) for $H^*$ under the constraint
$H^*(H=0,J=0)=0$ leads to
%For $\Delta q=q_{S0}-q_{G0}>0$,
%
\begin{align}\label{eq:1DH*}
H^*=&\frac{\omega_+^{2}\delta q_0}{{M_S\Omega}} -
\frac{\omega_+^{2}\delta q_0}{{M_S\Omega}}
\left\{\left(1-\frac{\delta q_0 - \delta q}{\delta q_0} \right)^2 \right.\\
& \left. \ \ \ +\frac{\nu_-^2 [(\delta \psi_0)^2-(\delta
\psi)^2]}{\omega_+^2 (\delta q_0)^2 } \rule{0mm}{6mm}
\right\}^{1/2}, \nonumber
\end{align}
where
%$\pm$ should be chosen in such a way .
$\delta q=q_{S}-q_{G}$, $\delta \psi=\psi_{S}-\psi_{G}$. Here we
have used the relations, $M_S\Omega(H-\beta\chi
J)=\omega_+^{2}(\delta q_0-\delta q)$ and $M_S \Omega \lambda \chi
J=\nu_-^{2}(\delta \psi_0 - \delta \psi)$ obtained from
Eqs.~(\ref{eq:1DdqG}), (\ref{eq:1DdpsiG}), (\ref{eq:1DdqS}), and
(\ref{eq:1DdpsiS}). Since $(\delta q_0 -\delta q)/\delta q_0 \ll
1$
%$(\Delta \psi_0 - \Delta \psi)/\Delta \psi_0 \ll 1$,
and $\{\nu_-^2 [(\delta \psi_0)^2-(\delta \psi)^2]\} / [\omega_+^2
(\delta q_0)^2]\ll 1$,
% is expected to be, in typical situations, of
%order one at best,
one can expand the curly bracket in
Eq.~(\ref{eq:1DH*}) to obtain
\begin{align}\label{eq:1DH*exp}
H^*(H,J) = & H-\beta' \chi J+\frac{M_S\Omega}{2\nu_-^2\delta
q_0}(\lambda\chi J)^2 \\
& -\frac{M_S\Omega}{2\omega_+^2\delta
q_0}\frac{\delta\psi_0}{\delta q_0}(\lambda\chi J)(H-\beta'\chi
J)\nonumber\\&+O(J^3).\nonumber
\end{align}
In case the $\psi$ dependence of the disorder potential energy is
negligible, $\beta'=\beta$, $\nu_-^{-2}=\delta \psi_0=0$, and the
effective field $H^*$ reduces to $H-\beta \chi J$. Then the points
in the $(H,J)$ plane with the same depinning rate will form
straight lines with the slope
$\beta \chi$.
% since $H=\beta\chi J+H^*$]}%$-\beta \chi$.

However in more general situations with the $\psi$ dependence of
the disorder potential energy, deviations from
this simple result will occur .
When $\nu_-^{-2}\neq 0$ but $\delta \psi_0=0$, the contour lines
of the equi-depinning rate will not be straight but instead form
parabolas in the $(H,J)$ plane with the coefficient of the
$J$-quadratic term proportional to $\nu_-^{-2}$. Note that this
quadratic contribution to $H^*$ is entirely due to the adiabatic
STT, while in case of $E_B$ , both the adiabatic and nonadiabatic
STTs can generate the $J$-quadratic contributions
[Eq.~(\ref{eq:1DEb})]. In this sense, $H^*$ allows clearer
separation between the adiabatic and nonadiabatic STT
contributions.
When $\nu_-^{-2}=0$ but $\delta \psi_0\neq 0$, the contour lines
of the equiv-depinning rate will form straight lines with the
modified slope, $\beta'\chi$. In this case, the value of $\beta'$
will fluctuate from potential wells to potential wells.

The above analysis provides experimental procedures to determine
whether or not the $\psi$ dependence of the disorder potential
energy is negligible in a given experiment; If the contour lines
of the equiv-depinning rate are not straight lines, $\nu_-^{-2}$
is not zero. If the slope of the lines tangential to the contour
lines at the points $(H,J=0)$ fluctuates from potential wells to
potential wells, $\delta \psi_0$ is not zero.
%
% When the $\psi$ dependence arises due to the
%position-to-position fluctuations of the magnetic anisotropy
%strength, $\nu_-^{-2}$ becomes nonzero. Then in addition to the
%simple contribution $H-\beta J$, there occurs an additional
%quadratic contribution [$\propto (\lambda \chi J)^2$] coming from
%the adiabatic STT. When the $\psi$ dependence arises due to the
%position-to-position fluctuations of the preferred magnetization
%direction, $\delta \psi_0$ becomes nonzero. Then $\beta'\neq
%\beta$ and there comes an additional contribution to $H^*$ from
%the second line in Eq.~(\ref{eq:1DH*exp}).

%%%%%%%%%%%%%%%%%%%%%%%%%%%%%%%%%%%%%%%%%%%%%%%%%%%%%%%%%%%%%%%%%
\subsection{Rashba spin-orbit coupling effects}\label{sec:1Drsoc}
%%%%%%%%%%%%%%%%%%%%%%%%%%%%%%%%%%%%%%%%%%%%%%%%%%%%%%%%%%%%%%%%%
%Rashba SOC effect
The special theory of relativity requires the coupling between the
spin and orbital degrees of freedom~\cite{gasi74}. Thus the
spin-orbit coupling (SOC) is ubiquitous. The strength of the SOC
however varies considerably from systems to systems. It is well
known~\cite{Winkler2003Book} that the SOC may be considerably
enhanced in systems with the broken inversion symmetry. The SOC in
this case is called the Rashba SOC (RSOC). Magnetic systems are
not exceptions and the RSOC develops in magnetic systems with the
broken inversion symmetry, as exemplified in a recent
experiment~\cite{Miron2010NatMat}.

Since the RSOC affects conduction electron spins and they in turn
interact with the local magnetization through the $s$-$d$ exchange
coupling, it also affects the local magnetization. It was
reported~\cite{moor08} that a high DW velocity can be achieved in
magnetic films with the broken inversion symmetry. In this
subsection, we discuss the RSOC effects on the DW depinning.

When the conduction electron spins are modified by the RSOC,
according to Ref.~\onlinecite{manc08}, the $s$-$d$ exchange coupling
generates an additional magnetic field acting on the local
magnetization. Though this is not a real magnetic field, it
behaves just like a real magnetic field as far as its effect on
the local magnetization is concerned.
When the inversion symmetry is broken along $\hat y$ direction and
the current is injected in $\hat x$
direction[Fig.~\ref{fig:rsocfield}(a)], this magnetic field
is~\cite{manc08}
\begin{equation}
\vec H_{\rm RSOC}=\frac{\alpha_R P}{\mu_B M_S}J(\hat x \times \hat
y),
\end{equation}
where $\alpha_R$ is the RSOC constant and $\mu_B$ is the Bohr
magneton\cite{manc08}. The direction of this field may or may not
be parallel to the real magnetic field applied to induce the DW
motion. When it is parallel, its effect is trivial since one just
needs to replace $H$ by $H+H_{\rm RSOC}$ in all equations
presented above. When it is not parallel, it may induce the
current-induced tilting angle jump at strong $H_{\rm RSOC}$,
 similarly to the chirality switching predicted
  for oblique magnetic field~\cite{Seo10APL}. For weak
$H_{\rm RSOC}$, the tilting angle jump is unlikely and a separate
analysis is required to understand its effect on the depinning.

As a representative example of nonparallel situations, we consider
a nanowire with the PMA~\cite{alva10, burr09, boul08, hein10,Kim09IEEE} along the
$\hat{y}$ direction. Then the external magnetic field $H$ (along
$\hat y$ direction) for the DW motion and ${\vec H}_{\rm RSOC}$
(along $\hat z$ direction) are mutually orthogonal.
%The DW depinning
%experiments used nanowires with the PMA.
%\textcolor[rgb]{1.00,0.00,0.00}{\textbf{[How about delete this
%sentence and move references to the first sentence of this
%paragraph?]}}
In PMA nanowires, two types of DWs can exist
depending on the width $w$ of the
nanowire\cite{swju08}[Fig.~\ref{fig:rsocfield}(b)].; When $w$ is
larger than a threshold value, a Bloch wall is energetically
preferred, and when $w$ is smaller than the threshold value, a
N\'eel wall is preferred.

%{\bf [This paragraph is new. Please read it carefully]}
%
One of primary effects of $\vec{H}_{\rm RSOC}$ is to modify the
$\psi$ dependence of $E(q,\psi)$ [Eq.~(\ref{eq:1DEn})], since the
energy of the system is minimized when the magnetization direction
at the center of the DW is parallel to ${\vec H}_{\rm RSOC}$.
Below we confine ourselves to the analysis of this additional
$\psi$ dependence, and ignore other effects of ${\vec H}_{\rm
RSOC}$. One example of the ignored effects is the $\psi$
dependence of the DW width $\lambda$. To be strict, $\lambda$
varies with $\psi$ even when ${\vec H}_{\rm
RSOC}=0$~\cite{Jung2007JM}, and nonzero ${\vec H}_{\rm RSOC}$
modifies the $\psi$ dependence of $\lambda$. This effect is
discussed in a recent experiment~\cite{Lavrusen2010Thesis}. For
${\vec H}_{\rm RSOC}=0$, it is commonly estimated that the $\psi$
dependence of $\lambda$ does not affect the DW motion
significantly for small $H$~\cite{naka03} and/or $J$.
We expect that at least for small ${\vec H}_{\rm RSOC}$, this
effect is still not important. Below we examine the small ${\vec
H}_{\rm RSOC}$ regime.

%i.e.
%the RSOC favors $\psi=0$ or $\pi$ depending on the sign of
%$\alpha_R$.
%When $w$ is larger than the threshold value, on the other hand, a
%N\'eel wall is energetically preferred and its magnetization
%direction at the DW center is parallel to $\vec H_R$.
%, $\vec H_R$ is perpendicular to the DW center magnetization.

%%%%%%%%%%%%%%%%%%%%%%%%%%%%%%%%%%%%%%%%%%%%%%
\subsubsection{Bloch DW} \label{sec:1DBlochDW}
%%%%%%%%%%%%%%%%%%%%%%%%%%%%%%%%%%%%%%%%%%%%%%
%RSOC effect on a Bloch wall
For a Bloch DW, the magnetization at the center of the DW points
along the $\hat{z}$ axis and we set $\psi=0$ for this direction.
Then $\vec H_{\rm RSOC}$ introduces an additional Zeeman energy
$E_{\rm RSOC}=-2M_S\Omega\lambda\chi J\tilde\alpha_R\cos\psi$ to
the system. Here, the dimensionless constant $\tilde
\alpha_R=(2\pi m\lambda/\hbar^2)\alpha_R$ measures the strength of
the RSOC. Then the total DW energy becomes
\begin{align}\label{eq:1DEnrsoc}
E(q,\psi)&=V(q,\psi)-2M_S\Omega q(H-\beta\chi
J)\nonumber\\&+2M_S\Omega\lambda\chi(\psi-\tilde\alpha_R\cos\psi)
J.
\end{align}
To calculate $E_B$ in the presence of $\tilde\alpha_R$, we need to
calculate the shifts of the saddle point and ground state configurations due to $H$ and
$J$ as we did in Sec.~\ref{sec:1DEb}. Since $E_{\rm RSOC}$ is
independent of $q$, it only affects the shifts of $\psi_S$ and
$\psi_G$. From $\delta E=0$, $\psi$ value of saddle(ground) point
$\psi_{S(G)}$ for finite $\tilde\alpha_R$ should satisfy
\begin{align}
\psi_{S(G)}=\psi_{S0(G0)}-\frac{M_S\Omega}{\nu_{S(G)}^2}\lambda\chi
J(1+\tilde\alpha_R\sin\psi_{S(G)}).
\end{align}
Since $\psi_{S(G)}-\psi_{S0(G0)} \ll 1$,
$\sin\psi_{S(G)}=\sin[\psi_{S0(G0)}+(\psi_{S(G)}-\psi_{S0(G0)})]$
may be Taylor expanded.
% as
% $\approx
%\sin\psi_{S0(G0)}+(\psi_{S(G)}-\psi_{S0(G0)})\cos\psi_{S0(G0)}$.
%{\bf [Check if we need to retain the quadratic term in this series
%expansion]} Then,
%\begin{align}\label{eq:1Ddpsirsoc}
%\psi_{S(G)}\approx\psi_{S0(G0)}-\frac{(M_S\Omega\lambda\chi
%J)(1+\tilde\alpha_R\sin\psi_{S0(G0)})}{\nu_{S(G)}^2+(M_S\Omega\lambda\chi
%J)\tilde\alpha_R\cos\psi_{S0(G0)}}.
%\end{align}
After some calculation, one then finds that up to ${\cal
O}(\tilde{\alpha}_R)$, $E_B$ is given by
%Substitution of
%Eq.~\eqref{eq:1Ddpsirsoc} to Eq.~\eqref{eq:1DEnrsoc} gives
%\begin{align}\label{eq:1DEbrsocBloch}
%E_B \approx & V_{0} -2\delta q_0 M_S\Omega (H-\beta\chi J)  \\
%& + 2
%[\delta\psi_0-\tilde\alpha_R(\cos\psi_{S0}-\cos\psi_{G0})]M_S\Omega(\lambda\chi J) \nonumber\\
%& +\frac{M_S^2\Omega^2}{\omega_+^{2}}(H-\beta\chi J)^2 \nonumber \\
%& -(M_S\Omega)^2
%\left[\frac{(1+\tilde\alpha_R\sin\psi_{S0})^{2}}{\nu_S^2
%+(M_S\Omega\lambda\chi J)\tilde\alpha_R\cos\psi_{S0}} \right. \nonumber\\
%&\ \ \ \ \ \ \ \ -\left.
%\frac{(1+\tilde\alpha_R\sin\psi_{G0})^{2}}{\nu_G^2+(M_S\Omega\lambda\chi
%J)\tilde\alpha_R\cos\psi_{G0}}\right](\lambda\chi J)^2. \nonumber
%%E_B&=V_{0}-2M_S\Omega(q_{S0}-q_{G0})(H-\beta\chi J)+M_S^2\Omega^2(\omega_S^{-2}+\omega_G^{-2})(H-\beta\chi J)^2 \nonumber\\ &+2M_S\Omega(\lambda\chi J)[(\psi_{S0}-\psi_{G0})-\tilde\alpha_R(\cos\psi_{S0}-\cos\psi_{G0})]\nonumber\\
%%&-(M_S\Omega\lambda\chi J)^2
%%[\frac{(1+\tilde\alpha_R\sin\psi_{S0})^2}{\nu_S^2+(M_S\Omega\lambda\chi J)\tilde\alpha_R\cos\psi_{S0}}-\frac{(1+\tilde\alpha_R\sin\psi_{G0})^2}{\nu_G^2+(M_S\Omega\lambda\chi J)\tilde\alpha_R\cos\psi_{G0}}].
%\end{align}
%Up to the ${\cal O}(\tilde\alpha_R)$, Eq.~\eqref{eq:1DEbrsocBloch}
%is {\bf [Check colored numbers in Eqs. (23) and (24)]}
\begin{align}
E_B \approx & V_{0} -2\delta q_0 M_S\Omega(H-\beta\chi J) \\
&+2[\delta\psi_0-\tilde\alpha_R(\cos\psi_{S0}-\cos\psi_{G0})]M_S\Omega(\lambda\chi J)\nonumber\\
&+{M_S^2\Omega^2}{\omega_+^{-2}}(H-\beta\chi J)^2
\nonumber\\
&-(M_S\Omega)^2 \left[\frac{1}{\nu_-^{2}}
+2\tilde\alpha_R\left(\frac{\sin\psi_{S0}}{\nu_S^{2}}
-\frac{\sin\psi_{G0}}{\nu_G^{2}}\right) \right. \nonumber\\
&\ \ \  \left. - (M_S\Omega\lambda\chi
J)\tilde\alpha_R\left(\frac{\cos\psi_{S0}}{\nu_S^{4}}-\frac{\cos\psi_{G0}}{\nu_G^{4}}\right)
 \right](\lambda\chi J)^2 \nonumber.
%E_B&\approx V_{0}-2M_S\Omega(q_{S0}-q_{G0})(H-\beta\chi J)+M_S^2\Omega^2(\omega_S^{-2}+\omega_G^{-2})(H-\beta\chi J)^2 \nonumber\\
%&+2M_S\Omega(\lambda\chi J)[(\psi_{S0}-\psi_{G0})-\tilde\alpha_R(\cos\psi_{S0}-\cos\psi_{G0})]\nonumber\\
%&-(M_S\Omega\lambda\chi J)^2
%[(\nu_S^{-2}-\nu_G^{-2})+\tilde\alpha_R(\nu_S^{-2}\sin\psi_{S0}
%-\nu_G^{-2}\sin\psi_{G0})\nonumber\\
%&+(M_S\Omega\lambda\chi J)\tilde\alpha_R(\nu_S^{-4}\cos\psi_{S0}-\nu_G^{-4}\cos\psi_{G0})].
\end{align}

Note that the nonadiabatic STT contribution to $E_B$ is not
modified by the RSOC. The RSOC effect modifies the adiabatic STT
contribution to $E_B$. Since the adiabatic STT contribution is
dependent on the $\psi$ dependence of the disorder potential, the
RSOC effect is also dependent on the $\psi$ dependence of the
disorder potential.
When the $\psi$-dependence of the disorder potential energy is
absent, $\nu_-^{-2}=\delta \psi_0=0$, one finds
\begin{align}\label{eq:ebbw}
E_B \approx & V_{0}-2\delta q_0 M_S\Omega (H-\beta\chi J) \\
&+{M_S^2\Omega^2}{\omega_+^{-2}}(H-\beta\chi J)^2. \nonumber
%-(M_S\Omega\lambda\chi J)^2
%\nu_-\nonumber\\&-\tilde\alpha_R(M_S\Omega\lambda\chi J)^2
%\nu_-[\sin\psi_{G0}+(M_S\Omega\lambda\chi J)\nu_+\cos\psi_{G0})],
\end{align}
Note that the result does not depend on $\tilde{\alpha}_R$.
When
%the strength ($\Omega \lambda K_d$) of the DW anisotropy
%energy strength fluctuates without the fluctuation in the
%preferred anisotropy direction,
%
%the $\psi$-dependence of the disorder potential energy arises
%from the position-by-position fluctuations of the saturation
%magnetization,
$\nu_-^{-2} \neq 0$ but $\delta \psi_0=0$ (also
$\psi_{G0}=\psi_{S0}=0$), one finds
\begin{align}
E_B \approx & V_{0}-2\delta q_0 M_S\Omega (H-\beta\chi J) \\
&+{M_S^2\Omega^2}{\omega_+^{-2}}(H-\beta\chi J)^2 \nonumber \\
& -{(M_S\Omega)^2}{\nu_-^{-2}}(\lambda\chi J)^2  \left(1
 - \tilde{\alpha}_R{M_S\Omega}{\nu_+^{-2}}   \lambda\chi J
 \right), \nonumber
%& \ \ \ \ \times \left[1+\tilde{\alpha}_R \left( 2 \sin \psi_{G0}
% \textcolor{red}{- \frac{M_S\Omega}{\nu_+^2}   \lambda\chi J
%\cos\psi_{G0} }\right) \right], \nonumber
%\nu_-[\sin\psi_{G0}+(M_S\Omega\lambda\chi J)\nu_+\cos\psi_{G0})],
\end{align}
where $\nu_+^{-2}\equiv \nu_S^{-2}+\nu_G^{-2}$.
Note that the leading effect of the RSOC is to introduce a
correction term proportional to $\tilde{\alpha}_R (\lambda \chi
J)^3$.
On the other hand, when
%the $\psi$-dependence of the disorder
%potential energy arises from the position-by-position fluctuations
%of the DW anisotropy direction,
$\nu_-^{-2}=0$ but $\delta \psi_0\neq 0$, one finds
\begin{align}
E_B \approx & V_{0} -2\delta q_0 M_S\Omega(H-\beta\chi J) \\
&+2[\delta\psi_0-\tilde\alpha_R(\cos\psi_{S0}-\cos\psi_{G0})]M_S\Omega(\lambda\chi J)\nonumber\\
&+{M_S^2\Omega^2}{\omega_+^{-2}}(H-\beta\chi J)^2
\nonumber\\
&-\tilde\alpha_R{(M_S\Omega)^2}{\nu_G^{-2}}(\lambda\chi J)^2
\left[\rule{0mm}{5mm}
2\left(\sin\psi_{S0}-\sin\psi_{G0}\right) \right. \nonumber\\
&\ \ \   \left. - {M_S\Omega}{\nu_G^{-2}}\lambda\chi
J\left(\cos\psi_{S0}-\cos\psi_{G0}\right)
 \right] \nonumber.
\end{align}
Again the RSOC modifies the adiabatic STT effect. Note that all
terms containing $\tilde \alpha_R$ are proportional to either
$\sin \psi_{G0}-\sin \psi_{S0}$ or $\cos \psi_{G0}-\cos
\psi_{S0}$, both of which vanish upon the average over many
potential wells.

%%%%%%%%%%%%%%%%%%%%%%%%%%%%%%%%%%%%%%%%%%%%%%
\subsubsection{N\'eel DW} \label{sec:1DNeelDW}
%%%%%%%%%%%%%%%%%%%%%%%%%%%%%%%%%%%%%%%%%%%%%%
%RSOC on a Neel wall
For a N\'eel DW, the magnetization at the center of the DW points
along the $\hat{x}$ axis and we set $\psi=0$ for this direction.
Then the Zeeman energy $E_{\rm RSOC}$ due to ${\vec H}_{\rm RSOC}$
becomes $E_{\rm RSOC}=-2M_S\Omega\lambda\chi
J\tilde\alpha_R\sin\psi$. Following the same procedure as above,
one obtains the energy barrier up to ${\cal O}(\tilde{\alpha}_R)$,
\begin{align}\label{eq:1DEbrsoc}
E_B \approx & V_{0}-2M_S\Omega\delta q(H-\beta\chi J) \\
&+2[\delta\psi_0-\tilde\alpha_R(\sin\psi_{S0}-\sin\psi_{G0})]M_S\Omega(\lambda\chi J)\nonumber\\
&+{M_S^2\Omega^2}{\omega_+^{-2}}(H-\beta\chi J)^2
\nonumber\\
&-(M_S\Omega)^2
\left[\frac{1}{\nu_-^{2}}-2\tilde\alpha_R\left(\frac{\cos\psi_{S0}}{\nu_S^{2}}
-\frac{\cos\psi_{G0}}{\nu_G^{2}}\right) \right. \nonumber\\
&\left. \ \  \ \ \ -(M_S\Omega\lambda\chi
J)\tilde\alpha_R\left(\frac{\sin\psi_{S0}}{\nu_S^{4}}
-\frac{\sin\psi_{G0}}{\nu_G^{4}}\right)\right](\lambda\chi J)^2.
\nonumber
%E_B&\approx V_{0}-2M_S\Omega(q_{S0}-q_{G0})(H-\beta\chi J)+M_S^2\Omega^2(\omega_S^{-2}+\omega_G^{-2})(H-\beta\chi J)^2 \nonumber\\
%&+2M_S\Omega(\lambda\chi J)[(\psi_{S0}-\psi_{G0})-\tilde\alpha_R(\sin\psi_{S0}-\sin\psi_{G0})]\nonumber\\
%&-(M_S\Omega\lambda\chi J)^2
%[(\nu_S^{-2}-\nu_G^{-2})-\tilde\alpha_R(\nu_S^{-2}\cos\psi_{S0}
%-\nu_G^{-2}\cos\psi_{G0})\nonumber\\
%&+(M_S\Omega\lambda\chi J)\tilde\alpha_R(\nu_S^{-4}\sin\psi_{S0}-\nu_G^{-4}\sin\psi_{G0})].
\end{align}

Similarly to the Bloch DW, the RSOC effect on the N\'eel DW
appears through the adiabatic STT contribution to $E_B$, and is
dependent on the $\psi$ dependence of the disorder potential
energy.
When the $\psi$-dependence of the disorder potential energy is
absent, $\nu^{-2}=\delta \psi_0=0$, one finds that Eq.~\eqref{eq:1DEbrsoc} becomes
equivalent to Eq.~\eqref{eq:ebbw}.
Note again, that the result does not depend on $\tilde\alpha_R$.
When $\nu^{-2}\neq 0$ but $\delta \psi_0=0$ (also
$\psi_{G0}=\psi_{S0}=0$), one finds
\begin{align}
E_B \approx & V_{0}-2M_S\Omega\delta q(H-\beta\chi J) \\
&+{M_S^2\Omega^2}{\omega_+^{-2}}(H-\beta\chi J)^2
\nonumber\\
&-{(M_S\Omega)^2}{\nu_-^{-2}}(1-2\tilde{\alpha}_R)(\lambda\chi J)^2. \nonumber\\
%\left[\frac{1}{\nu_-^{2}}-\tilde\alpha_R\left(\frac{\cos\psi_{S0}}{\nu_S^{2}}
%-\frac{\cos\psi_{G0}}{\nu_G^{2}}\right) \right](\lambda\chi J)^2 \nonumber\\
%&\left. \ \  \ \ \ +(M_S\Omega\lambda\chi
%J)\tilde\alpha_R\left(\frac{\sin\psi_{S0}}{\nu_S^{4}}
%-\frac{\sin\psi_{G0}}{\nu_G^{4}}\right)\right]. \nonumber
\end{align}
Note that the leading effect of the RSOC is to introduce a
correction term proportional to $\tilde{\alpha}_R (\lambda \chi
J)^2$.
On the other hand, when $\nu_-^{-2}=0$ but $\delta \psi_0\neq 0$,
one finds
\begin{align}\label{nweb3}
E_B \approx & V_{0}-2M_S\Omega\delta q(H-\beta\chi J) \\
&+2[\delta\psi_0-\tilde\alpha_R(\sin\psi_{S0}-\sin\psi_{G0})]M_S\Omega(\lambda\chi J)\nonumber\\
&+{M_S^2\Omega^2}{\omega_+^{-2}}(H-\beta\chi J)^2
\nonumber\\
&+\tilde{\alpha}_R {(M_S\Omega)^2}{\nu_G^{-2}}(\lambda\chi J)^2
\left[2(\cos\psi_{S0}-\cos\psi_{G0}) \right. \nonumber
\\
& \ \ \ \ \left. +{M_S\Omega}{\nu_G^{-2}}\lambda\chi J
\left(\sin\psi_{S0}-\sin\psi_{G0}\right)\right]. \nonumber
\end{align}
Again the RSOC modifies the adiabatic STT effect. Note that all
terms containing $\tilde{\alpha}_R$ in Eq.~\eqref{nweb3} vanish upon the averaging over
many potential wells.

%%%%%%%%%%%%%%%%%%%%%%%%%%%%%%%%%%%%%%
\section{DW creep in 2D}\label{sec:2D}
%%%%%%%%%%%%%%%%%%%%%%%%%%%%%%%%%%%%%%
%Description of a DW motion in 2D
When the thickness or the width of a magnetic nanowire is larger
than the collective length $L_{col}$, the system is not a 1D
problem any more. Here we assume that the width is sufficiently
larger than $L_{col}$ and the thickness is sufficiently smaller
than $L_{col}$, so that the system becomes a 2D problem. In the 2D
regime, the DW configuration can be described by two functions,
$q(z)$ and $\psi(z)$, where $z$ denotes the coordinates along the
nanowire width direction. In this Section, we examine the DW creep in this
2D regime. We find that the $\psi$-dependence of the disorder
potential energy again plays important roles, similarly to the 1D
case. Previous studies~\cite{duin08, luca09} of the DW creep motion
have ignored the $\psi$-dependence of the disorder potential
energy.

%%%%%%%%%%%%%%%%%%%%%%%%%%%%%%%%%%%%%%%%%%%%%%%%%%%%%%
\subsection{Effective energy barrier} \label{sec:2DEb}
%%%%%%%%%%%%%%%%%%%%%%%%%%%%%%%%%%%%%%%%%%%%%%%%%%%%%%

%%%%%%%%%%%%%%
\begin{figure}
\subfigure[]
{\includegraphics[width=0.15\textwidth]{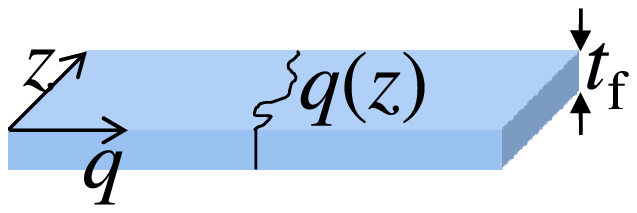}}\\
\subfigure[]
{\includegraphics[width=0.3\textwidth]{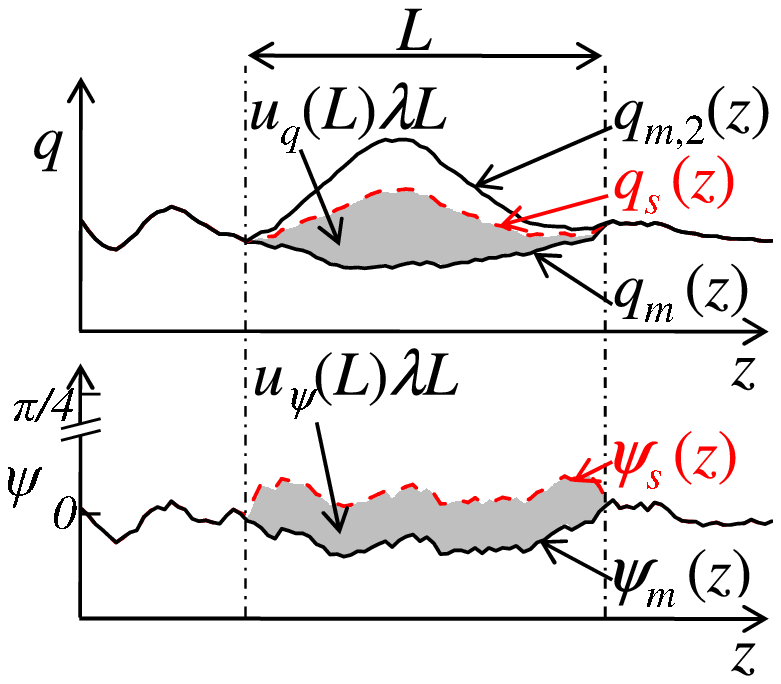}}
\caption{(Color online) (a) Schematic plot of the coordinates
system. (b) Schematic illustration of a DW segment of length $L$,
which makes a thermally-assisted transition from the original
local minimum configuration [$\{q_m(z)\}$,$\{\psi_m(z)\}$] to
another local minimum configuration
[$\{q_{m,2}(z)\}$,$\{\psi_{m,2}(z)\}$] through the saddle
configuration [$\{q_s(z)$\},\{$\psi_s(z)\}$]. The upper (lower)
panel shows the change of $\{q(z)\}$ ($\{\psi(z)\}$) during the
transition. The areas of the grey regions in upper and lower
panels correspond to $u_q(L)\lambda L$ and $u_\psi(L)\lambda L$,
respectively.}\label{fig:2Dmodel}
\end{figure}
%%%%%%%%%%%%

When the nanowire width $w$ is larger than $L_{col}$, an entire DW
line does not move simultaneously. Instead, a DW motion consists
of a segment-by-segment motion of DW segments of finite lengths.
In this situation, the thermally activated DW motion involves DW
segments of all possible segment lengths and the DW creep velocity
is governed by the bottleneck process with the largest energy
barrier~\cite{leme98,chau00}. Hence, the effective energy barrier
$E_B^{\rm creep}(H,J)$ for the DW creep motion, which determines
the DW velocity $v(H,J)\propto \exp[-E_B^{\rm creep}(H,J)/k_B T]$, becomes the
maximum value of $E_B(L)$ with respect to $L$, where $E_B(L)$
represents the effective energy barrier for a DW segment of length
$L$.

%V(q, psi), $V_q, V_psi$ separation
Figure~\ref{fig:2Dmodel}(a) depicts schematically the DW
configuration  in the 2D system.
%For
%a simplicity, $\vec R(z)=(q(z)/\lambda,\psi)$ is introduced.
According to Ref.~\onlinecite{duin08}, the effective energy
$E[\{q(z)\},\{\psi(z)\}]$ of a given DW configuration
$[\{q(z)\},\{\psi(z)\}]$ is given by
\begin{align}\label{eq:2DV}
E= \int\frac{dz}{\lambda} & \left\{  \frac{\tilde J}{2\hbar}\left[
\left(\frac{\partial q}{\lambda\partial
z}\right)^2+\left(\frac{\partial \psi}{\partial z}\right)^2\right]
\right.
\\
& -\frac{K_\perp}{4\hbar}\cos2\psi+V_{\rm dis}
\nonumber \\
&\left.-M_St_f(H-\beta\chi J)q+M_St_f\psi\lambda\chi
J\rule{0mm}{0.6cm}\right\}, \nonumber
\end{align}
where  $\tilde J$ measures the DW elasticity  and $K_\perp$
denotes the DW anisotropy\cite{swju08}. In Eq.~(\ref{eq:2DV}), the
first, second, and third terms represent the DW elastic energy,
the DW anisotropy energy, and the disorder potential energy,
respectively. The last term in Eq.~\eqref{eq:2DV} denotes the
effective energy due to the adiabatic STT and the second last term
denotes the combined effect of the Zeeman energy due to $H$ and
the effective energy due to the non-adiabatic STT.
One remark is in order.  As in the case of one-dimensional DW depinning
 in Sec.~\ref{sec:1DEb}, the effective energy $E$ in
Eq.~\eqref{eq:2DV} is a multi-valued function since
$E[\{q(z)\},\{\psi(z)\}]\neq E[\{q(z)\},\{\psi(z)+2\pi\}]$ while
two configurations $[\{q(z)\},\{\psi(z)\}]$ and
$[\{q(z)\},\{\psi(z)+2\pi\}]$ are physically identical.
Nevertheless this multi-valuedness problem does not cause any
ambiguity in the determination of $E_B(L)$ in Eq.~\eqref{eq:EL}
since $\psi$ is strictly confined to values much smaller than
$\pi/4$ in our study.

In general, $V_{\rm dis}$ will depend on both $q$ and $\psi$,
$V_{\rm dis}=V_{\rm dis}(q(z),\psi(z),z)$. Later we find that $\psi$ dependence can
generate interesting contributions, just as it did in the 1D
system. For definiteness of the illustration, we consider a
particular type of the $\psi$ dependence of $V_{\rm dis}$, arising
from the position-by-position fluctuation of $K_\perp$. Then the
fluctuating part $\delta K_\perp$ generates the contribution
$-(\delta K_\perp/4\hbar) \cos 2\psi$ to $V_{\rm dis}$. This
fluctuation can arise, for instance, from position-by-position
fluctuations of the saturation magnetization and nanowire
cross-section.
%In such a case, we may absorb the $\psi$ dependence
%of $V_{\rm dis}$ to $K_\perp$ and regard $V_{\rm dis}$ to be
%dependent only on $q$. Thus $K_\perp$ now becomes
%position-dependent and $K_\perp(q(z),z)\cos 2\psi(z)$ carries the
%$\psi$ dependence of the disorder potential.
For simplicity of the analysis, we ignore the fluctuating part
$\delta K_\perp$ for a while and consider it in the later part of
the analysis.

To evaluate $E$, it is useful to decompose it into two pieces
$E[\{q(z)\},\{\psi(z)\}]=E_q[\{q(z)\}]+E_\psi[\{\psi(z)\}]$, where
\begin{align}\label{eq:Vq}
E_q =  \int\frac{dz}{\lambda}  & \left[\frac{\tilde J}{2\hbar}
\left(\frac{\partial q}{\lambda\partial z}\right)^2 +V_{\rm dis}
\right.
\\
&  \left. -M_S t_f(H-\beta\chi J)q \rule{0mm}{5mm} \right] \nonumber\\
\label{eq:Vpsi} E_\psi = \int\frac{dz}{\lambda} &
\left[\frac{\tilde J}{2\hbar}\left(\frac{\partial\psi}{\partial
z}\right)^2-\frac{K_\perp}{4\hbar}\cos2\psi \right.
 \\
 & \left. +M_St_f\psi\lambda\chi J \rule{0mm}{5mm} \right].
 \nonumber
\end{align}
%%%%%

%V_q evaluation
As outlined above, to evaluate $E_B^{\rm creep}$, we
first need to calculate the effective energy barrier $E_B(L)$ that
a DW segment of finite length $L$ experiences~\cite{chau00}.
Suppose a DW segment of length $L$ ($0<z<L$) makes a
thermally-assisted transition from one local minimum configuration
$[\{q_m(z)\},\{\psi_m(z)\}]$ of the effective energy $E$ to
another local minimum configuration
$[\{q_{m,2}(z)\},\{\psi_{m,2}(z)\}]$ through the saddle point
configuration $[\{q_s(z)\},\{\psi_s(z)\}]$
[Fig.~\ref{fig:2Dmodel}(b)]. These three configurations differ in
the range $0<z<L$ but are essentially the same in the range $z<0$
and $z>L$ since only the DW segment of length $L$ makes a
thermally-assisted transition.
Then the energy barrier becomes
$E_B(L)=E[\{q_s(z)\},\{\psi_s(z)\}]-E[\{q_m(z)\},\{\psi_m(z)\}]$
and it can be decomposed into two pieces,
$E_q[\{q_s(z)\}]-E_q[\{q_m(z)\}]$ and
$E_\psi[\{\psi_s(z)\}]-E_\psi[\{\psi_m(z)\}]$.

%%%%%%%%%%%%%%%%%%%%%%%%%%%%%%%%%%%$$
\subsubsection{$q$ degree of freedom}
%%%%%%%%%%%%%%%%%%%%%%%%%%%%%%%%%%%$$
First, we evaluate $E_q[\{q_s(z)\}]-E_q[\{q_m(z)\}]$. The last
term in Eq.~\eqref{eq:Vq} gives rise to the contribution
$-M_St_f(H-\beta\chi J)u_q(L)L$, where
$u_q(L)=\int_0^L\frac{dz}{\lambda}[q_s(z)-q_m(z)]/L$[Fig.~\ref{fig:2Dmodel}(b)]
measures the typical value of the difference $q_s(z)-q_m(z)$ in
the region $0<z<L$. Since $q_s(z)-q_m(z)\approx 0$ for $z<0$ and
$z>L$, it is evident that $u_q(L)$ is a growing function of $L$
(Fig.~\ref{fig:2Dmodel}). According to the theory of interfaces in
disordered media\cite{tani09} where the disorder and the elastic
energy compete, $u_q(L)$ grows as a power law
$u_q(L)=u_{q0}(L/L_C)^\zeta$ where $u_{q0}$ is a characteristic
scaling constant, $\zeta$ is the wandering exponent and $L_C$ is
the Larkin length\cite{leme98,natt90,chau00}. For DWs formed in
metallic ferromagnetic films,
$\zeta=2/3$\cite{kjkim09,leme98,natt90,chau00,meta07}. To find out
the total contribution of all three terms in Eq.~\eqref{eq:Vq} to
$E_q[\{q_s(z)\}]-E_q[\{q_m(z)\}]$, we note that $E_q[\{q(z)\}]$
has the same form as the DW free energy for the purely
field-driven DW motion. This problem has been analyzed in
Ref.~\onlinecite{chau00}, and we borrow the calculation result of
Ref.~\onlinecite{chau00} to obtain the characteristic $L$ dependence of
$E_q[\{q_s(z)\}]-E_q[\{q_m(z)\}]$,
%%%%%
\begin{align}\label{eq:ELq}
& E_q[\{q_s(z)\}]-E_q[\{q_m(z)\}] \\
%\tilde V_q(q_s)-\tilde V_m(q_m)
& \cong\epsilon_{el}\frac{\{u_q(L)\}^2}{L} -M_St_f(H-\beta\chi
J)u_q(L)L, \nonumber
\end{align}
%%%%%
where the DW energy density $\epsilon_{el}=\tilde
J/2\hbar\lambda^2$. Here the first term includes the combined
contribution of the first two terms in Eq.~(\ref{eq:Vq}).

%%%%%%%%%%%%%%%%%%%%%%%%%%%%%%%%%%%$$%%%
\subsubsection{$\psi$ degree of freedom}
%%%%%%%%%%%%%%%%%%%%%%%%%%%%%%%%%%%$$%%%
%Evaluation of $V_psi$
Next, we evaluate $E_\psi[\{\psi_s(z)\}]-E_\psi[\{\psi_m(z)\}]$.
For a purely field-driven DW motion, $\psi$ degree of freedom does
not play any role for the DW creep motion if the system is in the
regime below the Walker breakdown (the same holds for the DW
depinning in 1D systems as well, see Sec.~\ref{sec:1D}). Then,
$E_\psi\{\psi_s(z)\}]-E_\psi[\{\psi_m(z)\}]$ is essentially
zero~\cite{meta07, kjkim09, leme98, natt90, chau00}. Thus the
central task is to determine the effect of $J$ on this difference.
An injection of $J$ induces an excitation of $\psi$. Since the DW
anisotropy ($-K_\perp \cos 2\psi$) favors $\psi=0$,  the growth of
$\psi$ is strongly suppressed when $K_\perp$ is large, which is
the conventional situations in metallic ferromagnetic systems (in
ferromagnetic semiconductors, $K_\perp$ is usually much smaller
and this may not be the case).  Then we can fairly assume that
$|\psi|<\pi/4$ during the DW motion. This assumption is valid even
when spatial fluctuations of $K_\perp$ exist, provided that
the magnitude of the $K_\perp$ fluctuations is sufficiently
smaller than the spatial average of $K_\perp$.
Under this assumption, $\cos 2\psi$ in Eq.~\eqref{eq:Vpsi} may be
Taylor expanded to obtain
%%%%%
\begin{align}\label{eq:Vpsiex}
E_\psi = \int\frac{dz}{\lambda} & \left[\frac{\tilde
J}{2\hbar}\left(\frac{\partial\psi}{\partial
z}\right)^2+\frac{K_\perp(q,z)}{2\hbar}\psi^2 \right.
 \\
 & \left. +M_St_f\psi\lambda\chi J \rule{0mm}{5mm} \right]
-\int\frac{dz}{\lambda}\frac{K_\perp(q,z)}{4\hbar}, \nonumber
%E_\psi&=\int\frac{dz}{\lambda} [\frac{\tilde
%J}{2\hbar}(\frac{\partial\psi}{\partial
%z})^2+\frac{K_\perp(q,z)}{2\hbar}\psi^2+M_St_f\psi\lambda\chi
%J]+\int_0^L\frac{dz}{\lambda}[-\frac{K_\perp(q,z)}{4\hbar}],
\end{align}
%%%%%
where the position dependence of $K_\perp$ is made manifest. The
last term of Eq.~(\ref{eq:Vpsiex}) can be absorbed to $V_{\rm
dis}(q,z)$ in $E_q$ to define a new effective disorder potential
$V_{\rm dis}^{\rm new}(q,z)$, $V_{\rm dis}^{\rm new}(q,z)\equiv
V_{\rm dis}(q,z)-K_\perp(q,z)/4\hbar$. As long as $K_\perp(q,z)$
has the same statistical properties as $V_{\rm dis}(q,z)$, the
$L$-dependence of $E_q[\{q_s(z)\}]-E_q[\{q_m(z)\}]$ in
Eq.~(\ref{eq:ELq}) remains essentially the same. Then we may
forget about the last term of $E_{\psi}$ in Eq.~\eqref{eq:Vpsiex}
and consider only the first three terms.

To obtain the $L$-dependence of
$E_\psi[\{\psi_s(z)\}]-E_\psi[\{\psi_m(z)\}]$, we first examine
characteristics of the saddle and minimum configurations. At these
configurations, $\delta E_\psi/\delta \psi=0$. Thus $\psi_s$ and
$\psi_m$ satisfy
%%%%%
\begin{align}
-\frac{\tilde J}{\hbar}\frac{\partial^2\psi}{\partial z^2}
+\frac{K_\perp(q,z)}{\hbar}\psi+M_St_f\lambda\chi
J=0,\label{eq:dVdpsi}
\end{align}
%%%%%
where $q$ in $K_\perp(q,z)$ denotes $q_m(z)$ and $q_s(z)$,
respectively, for $\psi=\psi_m(z)$ and $\psi_s(z)$. We analyze
Eq.~(\ref{eq:dVdpsi}) under the boundary condition,
$\psi_m(z)-\psi_s(z)\approx 0$ for $z<0$ and $z>L$.
Equation \eqref{eq:dVdpsi} is solved firstly for $J =0 $. Note that
Eq.~\eqref{eq:dVdpsi} has the same structure as the
Schr\"{o}dinger equation\cite{gasi74} $- (\hbar^2/2m)
\partial^2\Psi /\partial z^2+[U(z)-E]\Psi =0$ for a quantum
mechanical particle of the mass $m$ subject to the potential
energy $U(z)$ with the total energy $E$. In this analogy,
$K_\perp(q(z),z)/\hbar$ corresponds to the difference $U(z)-E$. In
quantum mechanics, it is well-known that when the total energy $E$
is smaller than the potential energy $U(z)$, the solution
$\Psi(z)$  is a sum of two exponentially growing functions; one
growing as $z$ becomes more positive and the other growing as $z$
becomes more negative. For both exponentially growing
functions, the rate of the exponential growth is roughly given by
$\sqrt{2m[U(z)-E]}/\hbar$ . This knowledge of the Schr\"{o}dinger
equation is directly applicable to Eq.~\eqref{eq:dVdpsi} since
$K_\perp $ stays positive for all $z$.
%when the position-dependent
%fluctuation of $K_\perp$ is small.
This analogy implies that small
change in $K_\perp$ within $0<z<L$ causes an exponentially large
change in $\psi$ at the boundaries $z=0$ and $L$ (large $L$ limit
is important for the DW creep motion). Combined with the boundary
condition, and recalling that Eq.~\eqref{eq:dVdpsi} is a linear
homogeneous equation, we then find that both $\psi_s$ and $\psi_m$
should be essentially zero. All other solutions of
Eq.~\eqref{eq:dVdpsi} cannot satisfy the boundary condition and
moreover violate the assumption $|\psi|\ll \pi/4$ due to their
exponential growth.

%\psi_s & \psi_m \propto J for J\neq 0
Next, one considers nonzero $J$. Since Eq.~\eqref{eq:dVdpsi}  is
then a linear inhomogeneous differential equation, its general
solution is a sum of the general homogeneous solution for $J=0$ and a particular solution for $J \neq 0$. Due to the
exponential growth, the general homogeneous solution should be set
to zero again and we need to find one nonzero particular solution,
which is consistent with the boundary condition and satisfies the
assumption $|\psi|\ll \pi/4$. While the exact form of the
particular solution is difficult to obtain, it is evident from the
structure of the linear differential equation
Eq.~\eqref{eq:dVdpsi} that the particular solution $\psi$ should
be proportional to $J$. Thus, $\psi_s\propto J$ and $\psi_m\propto
J$.
As for the $L$-dependence of $\psi_s$ and $\psi_m$, it is evident
that they cannot grow as a power law of $L$ since they are
strictly bounded below $\pi/4$. Thus we obtain $\psi_s \propto L^0
J$ and $\psi_m \propto L^0 J$. The proportionality factors of
$\psi_s$ and $\psi_m$ are different since $K_\perp(q,z)$ in
Eq.~(\ref{eq:dVdpsi}) amounts to $K_\perp(q_s(z),z)$ and
$K_\perp(q_m(z),z)$, and they are different.
%The proportionality factors depend on $K_\perp(q,z)$ and it is
%evident that they do not depend on $L$.
Then it is straightforward to verify that in the evaluation of
$E_\psi[\{\psi_s(z)\}]-E_\psi[\{\psi_m(z)\}]$, each of the first
three terms in Eq.~(\ref{eq:Vpsiex}) generates the contribution
proportional to $LJ^2$ for $\psi=\psi_s$ and $\psi=\psi_m$.
%Thus both $\tilde
%V_{\psi}({\psi}_s)$ and $\tilde V_{\psi}({\psi}_m)$ scale as
%$LJ^2$ and
Then the characteristic $L$ dependence of
$E_\psi[\{\psi_s(z)\}]-E_\psi[\{\psi_m(z)\}]$ may be expressed as
\begin{equation} \label{eq:ELpsi}
E_\psi[\{\psi_s(z)\}]-E_\psi[\{\psi_m(z)\}] \cong
M_St_f\lambda\chi J u_\psi(L)L,
\end{equation}
where
$u_\psi(L)=\int_0^L\frac{dz}{\lambda}[\psi_s(z)-\psi_m(z)]/L$
scales as $L^0$ with the proportionality constant scaling as
$J^1$. We remark that for
$E_\psi[\{\psi_s(z)\}]-E_\psi[\{\psi_m(z)\}]$ to have a nonzero
value, it is crucial to take into account the $q$-dependent
fluctuation of $K_\perp$. Without it, $\psi_s=\psi_m$ and
$u_\psi(L)=0$ since $K_\perp(q_s,z)=K_\perp(q_m,z)$ and both
$\psi_s$ and $\psi_m$ satisfy the exactly same equation
[Eq.~\eqref{eq:dVdpsi}].

One remark is in order. In Ref.~\onlinecite{duin08},
$E_\psi[\{\psi_s(z)\}]-E_\psi[\{\psi_m(z)\}]$ was evaluated to be
proportional to $JL$, which is different from our evaluation
result $J^2 L$. This difference stems from the fact that the
thermally-activated transition process considered in
Ref.~\onlinecite{duin08} is qualitatively different from the transition
process considered in our work; While $|\psi|$ is assumed to
remain smaller than $\pi /4$ for the transition process considered
in our work, it is assumed in Ref.~\onlinecite{duin08} that $\psi$ jumps
by $\sim\pi$  for each transition process. Such transition with
the jump of $\psi$ by $\sim\pi$ may be relevant for a DW motion in
ferromagnetic semiconductors where the magnetic anisotropy is much
smaller.

%%%%%%%%%%%%%%%%%%%%%%%%%%%
\subsection{Creep velocity}
%%%%%%%%%%%%%%%%%%%%%%%%%%%
%E(L)
The DW velocity $v(H,J)$ in the creep regime is given by $v\propto
\exp(-E_B^{\rm creep}/k_B T)$, where $E_B^{\rm creep}$ for given
$H$ and $J$ is the maximum value of $E_B(L)$ with respect to $L$.
By combining Eqs.~(\ref{eq:ELq}) and (\ref{eq:ELpsi}), we obtain
the effective energy barrier $E_B(L)$ for the DW segment of length
$L$. Its $L$, $J$ and $H$ dependence can be summarized as
\begin{align}\label{eq:EL}
E_B(L)=&\epsilon_{el}\{u_q(L)\}^2L^{-1} \\
& -M_St_f(H-\beta\chi J)u_q(L)L \nonumber \\
& +M_St_f\lambda\chi J u_\psi(L)L, \nonumber
\end{align}
%%%%%
where $u_q(L)=u_{q0}(L/L_C)^\zeta$ and $u_\psi(L)=u_{\psi 0} L^0
J$. Substituting these relations into Eq.~(\ref{eq:EL}) leads to
%%%%%
\begin{align}\label{eq:ell}
E_B(L)=&\epsilon_{el}\frac{u_{q0}^2}{L_C^{2\zeta}}L^{2\zeta-1}
-M_St_f(H-\beta\chi J)\frac{u_{q0}}{L_C^{\zeta}}L^{\zeta+1}
\\
& +M_St_f\lambda\chi J^2 u_{\psi 0}L. \nonumber
\end{align}
%%%%%

For metallic ferromagnets\cite{meta07,
leme98,kjkim09,natt90,chau00}  with $\zeta =2/3$,
Eq.~\eqref{eq:ell} becomes
%%%%%
\begin{equation}
E_B(L)=AL^{1/3}-BL^{5/3}+CL\label{eq:ela}
\end{equation}
%%%%%
where $A=\epsilon_{el}u_{q0}^2 L_C^{-4/3}$, $B=M_St_f(H-  \beta
\chi J)u_{q0}L_C^{-2/3}$, and $C=M_St_f\lambda \chi u_{\psi
0}J^2$. The maximum energy barrier $E_B^{\rm creep}$ is then
determined by $E_B^{\rm creep}=E_B(L_{col})$, where the collective
length $L_{col}$ satisfies $\partial E_B/\partial L|_{L_{col}}=0$.
From Eq.~\eqref{eq:ela}, the collective length\cite{kjkim09}
$L_{col}$ is given by
%%%%%
\begin{equation}
L_{col}=\left(\frac{-3C+\sqrt{9C^2+20AB}}{2A}\right)^{-3/2},
\end{equation}
%%%%%
and $E_B^{\rm creep}$ is written as
\begin{equation}
E_B^{\rm
creep}=\frac{2}{5}(2A)^{3/2}\frac{(-2C+\sqrt{9C^2+20AB})}{(-3C+\sqrt{9C^2+20AB})^{3/2}}.
\label{eq:2DEM}
\end{equation}

%%%%%%%%%%%%%%%%%%%%%%%%%%%%%%%%%%%%%%%%%%%%%%%%%%%%%%%%
\subsubsection{Effective magnetic field}\label{sec:2DH*}
%%%%%%%%%%%%%%%%%%%%%%%%%%%%%%%%%%%%%%%%%%%%%%%%%%%%%%%%
The effective magnetic field $H^*(H,J)$ for the DW creep motion is
defined by the relation $v(H,J)=v(H^*,0)$ with the constraint
$H^*(H,0)=H$.  The effective magnetic field $H^*$ provides a
convenient way to express the result for $v(H,J)$; Recalling that
the DW velocity for the purely field-driven DW motion is
given~\cite{chau00} by $v(H,0)=v_0 \exp(-\kappa H^{-\mu}/k_B T)$,
the DW velocity for general $H$ and $J$ can be expressed as
\begin{equation} \label{eq:2D creep formula}
v(H,J)=v_0 \exp\left\{ -\frac{\kappa [H^*(H,J)]^{-\mu}}{k_B T}
\right\},
\end{equation}
where $\kappa$ is a constant independent of $H$ and $J$. Thus the
evaluation of $H^*(H,J)$ amounts to the evaluation of $v(H,J)$.
$H^*(H,J)$ also determines contour lines of equal DW velocity in
the $(H,J)$ plane.

Since $v(H,J)$ is determined by $E_B^{\rm creep}(H,J)$, $H^*(H,J)$
can be calculated from $E_B^{\rm creep}(H,J)=E_B^{\rm
creep}(H^*,0)$. We define $D=M_St_fu_{q0}L_C^{-2/3}$ and
$\epsilon= \beta \chi$. Then Eq.~(\ref{eq:2DEM}) can be expressed
as
% Replacing $B$ by $D(H-\epsilon J)$  where
%$D=M_St_fu_{q0}L_C^{-2/3}$ and $\epsilon= \beta \chi$,
%Eq.~\eqref{eq:2DEM} becomes
%\begin{align}\label{eq:2DEb}
%E_M=\frac{2}{5}\frac{(2A)^{3/2}}{(20AD)^{1/4}}\frac{(-\frac{2C}{\sqrt{20AD}}+\sqrt{\frac{9C^2}{20AD}+(H-\epsilon J)})}{(-\frac{3C}{\sqrt{20AD}}+\sqrt{\frac{9C^2}{20AD}+(H-\epsilon J)})^{3/2}}
%\end{align}
%
%To evaluate the effective total magnetic field in two-dimensional
%DW creep,  we can rewrite Eq.~\eqref{eq:2DEb} as
\begin{align}
E_B^{\rm creep}=\frac{2}{5}(2A)^{3/2}(20AD)^{- 1/4}[F(H,J)]^{-
1/4},\label{eq:2DEb2}
\end{align}
where
\begin{equation}
%F(H,J)=\frac{ \displaystyle \left[-\frac{3\eta
%J^2}{10}+\sqrt{\left(\frac{3\eta J^2}{10}\right)^2+(H-\epsilon
%J)}\right]^6}{\displaystyle \left[-\frac{\eta
%J^2}{5}+\sqrt{\left(\frac{3\eta J^2}{10}\right)^2+(H-\epsilon
%J)}\right]^4},\label{eq:F}
F(H,J)=\frac{\left[-3\eta J^2/10+\sqrt{(3\eta
J^2/10)^2+(H-\epsilon J)}\right]^6}{\left[-\eta J^2/5+\sqrt{(3\eta
J^2/10)^2+(H-\epsilon J)}\right]^4},\label{eq:F}
\end{equation}
where
%$3\eta J^2/10=3C/(20AD)^{1/2}$ and
$\eta=u_{\psi 0}\lambda
L_C\chi(5M_St_f/\epsilon_{el}u_{q0}^3)^{1/2}$. It can be easily
verified that $F(H,J=0)=H$. Since the constants $A$ and $D$ are
independent of $H$ and $J$, $F(H,J)$ itself is the effective
field, $H^*=F(H,J)$. One also finds that $\kappa$ in
Eq.~(\ref{eq:2D creep formula}) is given by
$\kappa=(2/5)(2A)^{3/2}(20AD)^{- 1/4}$. In the limit
$H,J\rightarrow 0$,
% When $H-\epsilon J$ is
%larger than $3\eta J^2/10$,
we expand $F(H,J)$ to obtain
\begin{equation} \label{eq:2DH*}
H^*(H,J)= H-\epsilon J-\eta J^2\sqrt{H-\epsilon
J}+\frac{2}{5}\left(\eta J^2\right)^2+{\cal O}\left(J^6\right).
\end{equation}
Again, as the DW depinning in 1D systems (Sec.~\ref{sec:1DH*}),
the non-adiabatic STT ($\epsilon J$) acts in the exactly same way
as the magnetic field ($H$). The adiabatic STT contribution ($\eta
J^2$), however, introduces the nonlinearity to $H^*$ and thus
plays a qualitatively different role from the magnetic field for
the creep motion.
%Since the adiabatic STT gives a $O(J^2\sqrt{H})$
%to the total effective magnetic field, we can calculate the
%significance of the adiabatic STT on the creep motion by measuring
%the nonlinear contribution of $J$ to the effective total magnetic
%field $H^*$.
If an experiment is performed for sufficiently small $H$ and $J$,
so that the nonlinear contributions in Eq.~(\ref{eq:2DH*}) are
negligible, the creep motion will follow a simple scaling
behavior, $v(H,J)=v_0 \exp [ -\kappa (H-\epsilon J)^{-\mu}/k_B T]$
with $\mu=1/4$. However if $H$ and $J$ are not sufficiently small,
the nonlinear contributions in Eq.~(\ref{eq:2DH*}) introduce
deviations from the simple scaling behavior and should be taken
into account in an experimental analysis.

%comparision w/ Duine 08
%One last remark is in order.  In the $H\rightarrow 0$ limit, the
%energy barrier $E_B(L)$ in Eq.~\eqref{eq:EL} does not agree with
%the energy barrier function obtained in Ref.~\cite{duin08} [Eq.
%(17) in Ref.~\cite{duin08}] for the purely current-driven DW
%motion; While the third term of $E_B(L)$ in Eq.~\eqref{eq:ell} is
%proportional to $J^2$, the corresponding term in
%Ref.~\cite{duin08} is proportional to $J$. This difference stems
%from the fact that the thermally-activated transition process
%considered in Ref.~\cite{duin08} is qualitatively different from
%the transition process considered in our work; While $|\psi|$ is
%assumed to remain smaller than $\pi /4$ for the transition process
%considered in our work, it is assumed in Ref.~\cite{duin08} that
%$\psi$ jumps by $\pi$  for each transition process. Such
%transition may be relevant for a DW motion in ferromagnetic
%semiconductors where the magnetic anisotropy is much smaller.

%%%%%%%%%%%%%%%%%%%%%%%%%%%%%%%%%%%%%%%%%%%%%%%%%%%%%%%%%%%%%%%%%
\subsection{Rashba spin-orbit coupling effects}\label{sec:2Drsoc}
%%%%%%%%%%%%%%%%%%%%%%%%%%%%%%%%%%%%%%%%%%%%%%%%%%%%%%%%%%%%%%%%%
%$H_R$
The RSOC is generated when the inversion symmetry is
broken~\cite{Winkler2003Book}. When a current flows in  a nanowire
with the broken inversion symmetry, the magnetization feels as if
there is an additional magnetic field $\vec H_{\rm RSOC}$, of which
magnitude is proportional to $J$~\cite{manc08}. We consider
the case where the inversion symmetry along the $\hat{y}$ axis is
broken and the current flows along the $\hat{x}$ direction
(parallel to the DW motion direction). Then $\vec H_{\rm RSOC}$ is
along the $\hat{z}$ direction.
When the RSOC is strong, it may modify the nature of the DW motion
qualitatively. But when the RSOC is weak, its effect may be
accounted for perturbatively. Below we assume the RSOC
to be weak. Then its effect can be calculated in a way similar to
the 1D case discussed in Sec.~\ref{sec:1Drsoc}. Again the RSOC
effect varies depending on the magnetic anisotropy and the DW
structure. We confine ourselves to nanowires with the
PMA and consider two types of DW
structure; Bloch DW and N\'eel DW.

%%%%%%%%%%%%%%%%%%%%%%%%%%%%%%%%%%%%%%%%%%%%%%
\subsubsection{Bloch DW} \label{sec:2DBlochDW}
%%%%%%%%%%%%%%%%%%%%%%%%%%%%%%%%%%%%%%%%%%%%%%
The magnetization direction at the center of the Bloch DW points
along the $\hat z$ direction. In the convention where $\psi=0$ for the
Bloch DW, an additional Zeeman energy $E_{\rm RSOC}$ due to the
RSOC effect becomes
\begin{equation} \label{eq:E RSOC Bloch}
E_{\rm RSOC}=-2\int \frac{dz}{\lambda}M_st_f\lambda\chi
J\tilde\alpha_R\cos\psi,
\end{equation}
where $\tilde{\alpha}_R$ is the dimensionless RSOC coefficient
$\tilde{\alpha}_R=(2\pi m \lambda/\hbar^2)\alpha_R$. Depending on
the sign of $\tilde{\alpha}_R$, the RSOC may enhance or suppress
possible deviations from $\psi=0$.

Since $E_{\rm RSOC}$ depends only on $\psi$, it may be included in
$E_{\psi}$. Then Eq.~\eqref{eq:Vpsi} is modified to
\begin{align}\label{eq:Vpsirsoc}
E_{\psi}=\int\frac{dz}{\lambda} & \left[\frac{\tilde J}{2\hbar}
\left(\frac{\partial\psi}{\partial z}\right)^2-\frac{K_\perp}{4\hbar}\cos2\psi \right. \\
& \left. +M_St_f\lambda\chi
J(\psi-2\tilde\alpha_R\cos\psi)\rule{0mm}{6mm} \right]. \nonumber
\end{align}
For $|\psi|\ll\pi/4$, it reduces to
\begin{align} \label{eq:Vpsirsocex}
E_{\psi}=&\int\frac{dz}{\lambda} \left[\frac{\tilde J}{2\hbar}
\left(\frac{\partial\psi}{\partial z}\right)^2
+\frac{K_\perp(q,z)}{2\hbar}\psi^2 \right. \\
&  \ \ \ \ \ \ \ \ \left. +M_St_f\lambda\chi
J(\psi+\tilde\alpha_R\psi^2) \rule{0mm}{5mm} \right]
\nonumber\\
&+\int\frac{dz}{\lambda}
\left[-\frac{K_\perp(q,z)}{4\hbar}-2M_St_f\lambda\chi
J\tilde\alpha_R\right]. \nonumber
\end{align}
The second integral of Eq.~\eqref{eq:Vpsirsocex} can be absorbed
to $V_{\rm dis}$ in $E_q$ in Eq.~\eqref{eq:Vq}, and we may
concentrate on the first integral of Eq.~\eqref{eq:Vpsirsocex}.
Note that the contribution from the RSOC ($\propto J
\tilde{\alpha}_R \psi^2$) has the same structure as the DW
anisotropy contribution ($\propto K_\perp \psi^2$). Thus the main
effect of the RSOC is to renormalize $K_\perp$ to $\xi
K_\perp$, where $\xi=1+2\hbar M_S t_f \lambda \chi J
\tilde{\alpha}_R/K_\perp$. Since $\xi-1 \propto
\tilde{\alpha}_R J$, it is safe to assume $\xi-1 \ll 1$ in the
creep regime where $J$ is small.

For $\tilde{\alpha}_R=0$, it has been demonstrated
[Eq.~(\ref{eq:ELpsi})] that
$E_\psi[\{\psi_s(z)\}]-E_\psi[\{\psi_m(z)\}] \cong
M_St_f\lambda\chi J u_\psi(L)L$, where $u_\psi=u_{\psi 0}J$. For
nonzero $\tilde{\alpha}_R$, the RSOC effect will appear through
the renormalization of $u_{\psi 0}$. It is reasonable to expect
that the renormalized $u_{\psi 0}$ depends on $\xi-1$ in a
nonsingular way. Then we may Taylor expand $u_{\psi 0}$ with
$\xi -1$ as a small variable, and express the renormalized
$u_{\psi 0}$ as $u_{\psi 0}[1+\gamma_R J+{\cal O}(J^2)]$. Although
the exact evaluation of $\gamma_R$ is difficult, it is evident
that it should be proportional to $\tilde{\alpha}_R$.

%
%At local minimum and saddle configurations, $\delta \tilde
%V_{\psi}/\delta\psi=0$. Then, $\psi_m$ and $\psi_s$ should satisfy
%\begin{align}
%-\frac{\tilde J}{\hbar}\frac{\partial^2\psi}{\partial z^2}
%+(\frac{K_\perp(q,z)}{\hbar}-2\tilde\alpha_R M_St_f\lambda\chi
%J)\psi+M_St_f\lambda\chi J=0.\label{eq:dVdpsirsoc}
%\end{align}
%Since the DW structure is not stable anymore if the additional
%energy due to the RSOC($\propto\alpha_R J$) is larger than the
%magnetic anisotropy energy($\propto K_\perp$), the coefficient of
%$\psi$ in
%Eq.~\eqref{eq:dVdpsirsoc}(${K_\perp(q,z)}/{\hbar}-2\tilde\alpha_R
%M_St_f\lambda\chi J$) is positive for all $z$. Then, for the same
%reason as discussed in Sec.~\ref{sec:2DEb}, the general
%homogeneous solution of Eq.~\eqref{eq:dVdpsirsoc} is
%$\psi_s=\psi_m=0$. From the structure of
%Eq.~\eqref{eq:dVdpsirsoc}, the particular solution should be
%proportional to $J/(K_\perp-2\hbar M_St_f\lambda\chi\tilde\alpha_R
%J)\approx (J/K_\perp)[1+2(\hbar
%M_St_f\lambda\chi/K_\perp)\tilde\alpha_R J]$. Then, each term of
%the first integral in Eq.~\eqref{eq:Vpsirsocex} should be scaled
%as $LJ^2[1+\gamma_R J+O(\tilde\alpha_R^2J^2)]$ where $\gamma_R$ is
%a coefficient proportional to $\tilde\alpha_R$. Thus $\tilde
%V_\psi(\psi_s)-\tilde V_\psi(\psi_m)= M_St_f\lambda\chi u_{\psi 0}
%J^2(1+\gamma_R J) L$ up to the first order of $\tilde\alpha_R J$.
%

In the presence of the RSOC, the energy barrier $E_B(L)$ in
Eq.~(\ref{eq:ell}) is modified to
%for a DW segment with length $L$ then becomes
\begin{align}\label{eq:ellrsoc}
E_B(L)=&\epsilon_{el}\frac{u_{q0}^2}{L_C^{2\zeta}}L^{2\zeta-1}
-M_St_f(H-\beta\chi J)\frac{u_{q0}}{L_C^{\zeta}}L^{\zeta+1}
\\
& +M_St_f\lambda\chi J^2 u_{\psi 0}(1+\gamma_R J) L. \nonumber
\end{align}
Since Eq.~\eqref{eq:ellrsoc} has the same structure as
Eq.~\eqref{eq:ell} except that the last term of
Eq.~\eqref{eq:ellrsoc} is multiplied by the extra factor
$(1+\gamma_R J)$, the energy barrier $E_B^{\rm creep}$ for the
creep motion can be obtained straightforwardly from Eq.~\eqref{eq:2DEb2}.
For metallic ferromagnets with $\zeta=2/3$, the effective field for the Bloch DW in the presence of the RSOC is given by the equation identical to Eq.~\eqref{eq:2DEb2} except $\eta$ is now replaced by $\eta(1+\gamma_R J)$. The leading correction due to the RSOC($\propto\gamma_R$) appears in terms of cubic and higher orders of $J$ and thus, we conclude that the RSOC does not modify the DW creep motion qualitatively in small $J$ regime.
%For metallic ferromagnets with $\zeta=2/3$,
%one obtains
%\begin{align}
%E_B^{\rm creep}=\frac{2}{5}(2A)^{3/2}(20AD)^{- 1/4}\{F'(H,J)\}^{-
%1/4},
%\end{align}
%where
%\begin{equation}
%F'(H,J)=\frac{\left[-3\eta' J^2/10+\sqrt{(3\eta'
%J^2/10)^2+(H-\epsilon J)}\right]^6}{\left[-\eta'
%J^2/5+\sqrt{(3\eta' J^2/10)^2+(H-\epsilon J)}\right]^4}.
%F(H,J)=\frac{(-3\eta J^2(1+\gamma_R J)/10+\sqrt{(3\eta
%J^2(1+\gamma_R J)/10)^2+(H-\epsilon J)})^6}{(-\eta J^2(1+\gamma_R
%J)/5+\sqrt{(3\eta J^2(1+\gamma_R J)/10)^2+(H-\epsilon
%J)})^4}.
%\end{equation}
%Here $\eta'=\eta(1+\gamma_R J)$. The definitions of $A$, $D$,
%$\eta$ and $\epsilon$ are the same as those in Eq.~\eqref{eq:2DEb2}.
%After expansion, the effective field for a Bloch DW in the
%presence of the RSOC is
%\begin{equation}
%H^*(H,J) = H-\epsilon J-\eta' J^2\sqrt{H-\epsilon J}+{\cal
%O}(J^4).
%\end{equation}
%Thus the leading correction due to the RSOC effect is to replace
%$\eta$ in Eq.~(\ref{eq:2DH*}) by $\eta'=\eta(1+\gamma_R J)$, where
%$\gamma_R$ is proportional to $\tilde{\alpha}_R$.
%%
%Since the correction by nonzero $\tilde{\alpha}_R$ appears in
%rather high order terms in $J$, we conclude that the RSOC does not
%modify the DW creep motion qualitatively.

%%%%%%%%%%%%%%%%%%%%%%%%%%%%%%%%%%%%%%%%%%%%%%
\subsubsection{N\'eel DW} \label{sec:2DNeelDW}
%%%%%%%%%%%%%%%%%%%%%%%%%%%%%%%%%%%%%%%%%%%%%%
The magnetization direction at the center of the N\'eel DW points
along the nanowire direction ($\hat{x}$ direction). In the
convention where $\psi=0$ for this direction, an additional Zeeman
energy $E_{\rm RSOC}$ due to the RSOC effect becomes
\begin{equation}
E_{\rm RSOC}=-2\int \frac{dz}{\lambda}M_st_f\lambda\chi
J\tilde\alpha_R\sin\psi.
\end{equation}
Note that this equation differs from Eq.~(\ref{eq:E RSOC Bloch})
($\sin \psi$ vs. $\cos \psi$) since $\psi=0$ represents the
different directions ($\hat{z}$ vs. $\hat{x}$) in the two cases.
For $|\psi|\ll \pi/4$, $E_\psi$ in Eq.~(\ref{eq:Vpsi}) is modified
to
\begin{align}\label{eq:Vpsiarexn}
\tilde V_{\psi}(\psi)
%\int_0^L\frac{dz}{\lambda}[\frac{\tilde J}{2\hbar}(\frac{\partial\psi}{\partial z})^2-\frac{K_\perp}{4\hbar}\cos2\psi\nonumber\\
%&+M_St_f\lambda\chi J(\psi+2\tilde\alpha_R\sin\psi)]\label{eq:Vpsiarn}\\
\approx & \int\frac{dz}{\lambda}\left[\frac{\tilde J}{2\hbar}
\left(\frac{\partial\psi}{\partial z}\right)^2+\frac{K_\perp(q,z)}{2\hbar}\psi^2 \right. \\
&\ \ \ \ \ \ \ \ +\left. M_St_f\lambda\chi
J(1-2\tilde\alpha_R)\psi \rule{0mm}{5mm}\right]
\nonumber\\
&-\int\frac{dz}{\lambda}\frac{K_\perp(q,z)}{4\hbar}. \nonumber
\end{align}
Note that $\tilde{\alpha}_R$ appears only in the second line, which
accounts for the adiabatic STT effect. It is then evident that the
RSOC renormalizes the adiabatic STT effect by the renormalization
factor $(1-2\tilde{\alpha}_R)$.

With this knowledge, the energy barrier $E_B^{\rm creep}$ can be
obtained in a straightforward way.  For metallic ferromagnets with
$\zeta=2/3$, the effective field for the N\'eel DW is given by the equation identical to Eq.~\eqref{eq:2DEb2} except replacing $\eta$ by $\eta(1+2\tilde\alpha_R)$.
Note that the correction by nonzero $\tilde{\alpha}_R$ again
appears in rather high order terms in $J$. Thus we conclude that
the RSOC does not modify the creep motion of the N\'eel DW
qualitatively.

\section{conclusion} \label{sec:conclusion}
%%%%%%%%%%%%%%%%%%%%%%%%%%%%%%%%%%%%%%%%%%%
Magnetic DW motion in a nanowire was examined in the weak driving
force regime with particular attention to metallic ferromagnets,
where the DW anisotropy is very large. Effects of the magnetic
field, the adiabatic STT, and the nonadiabatic STT on the DW
motion were examined under the assumption that the amplitude of
the tilting angle dynamics is much smaller than $2\pi$. To be more
specific, we examined two phenomena, the DW depinning from a
single potential well in 1D systems, and the DW creep motion
through a disordered potential profile in 2D systems.

The analysis on the 1D depinning becomes relevant when both the
width and the thickness of a nanowire are smaller than the
collective length $L_{\rm col}$. The nonadiabatic STT has the same
effect as the magnetic field, and together, they generate the
leading order contribution to the depinning rate. We found that
the way that the adiabatic STT affects the DW depinning depends on
the nature of disorders. In particular, it was demonstrated that
in certain types of disorders, the conventional ways to determine
the nonadiabaticity parameter $\beta$ result in incorrect values.
Possible ways to avoid the incorrect evaluation have been
proposed.

The analysis on the 2D creep motion becomes relevant when the
width of a nanowire is larger than $L_{\rm col}$ while the
thickness remains smaller than $L_{\rm col}$.
A thermally-assisted DW velocity is determined  by the energy
barrier $E_B^{\rm creep}$ between two spatially adjacent local
minimum configurations in the DW energy profile. The contribution
of the non-adiabatic STT ($\propto\beta J$) to $E_B^{\rm creep}$
is the same as that of the magnetic field. The role of the
adiabatic STT, however, is qualitatively different from those of
the non-adiabatic STT and the magnetic field. Efficiencies of
driving forces (magnetic field and current) are described in terms
of the total effective magnetic field.
Both the magnetic field and the non-adiabatic STT generate linear
contributions to the total effective magnetic field, implying that
the purely field-driven and purely current-driven DW creep motions
belong to the same universality class.
The adiabatic STT, on the other hand, generates $J$-quadratic or
higher order contributions to the total effective magnetic field, and
thus its contributions constitute the next leading order
contributions. Although these contributions are irrelevant in the
vanishing driving force limit, their effects may need to be taken
into account in practical scaling analysis since experiments are
always carried out at small but finite driving force strength.
%When the adiabatic STT is not negligible or dominates the
%creep motion, however, the creep dynamics is qualitatively
%different from the magnetic-field driven creep motion.
%In both cases, by analyzing the linear contribution of $J$ on the
%effective total magnetic field, the magnitude of the non-adiabatic
%STT in the system can be calculated. The size of the adiabatic STT
%can be measured from the nonlinear contribution of $J$.

Effects of the Rashba spin-orbit coupling (RSOC)  on the DW
depinning in 1D systems and on the DW creep in 2D systems are also
discussed. For a Bloch wall in a nanowire with the PMA,
%where the effective magnetic field
%due to the RSOC ($\vec H_R$) is perpendicular to the magnetic hard
%axis,
the RSOC effect appears in terms of cubic and higher orders of $J$
in the effective energy barrier. For a N\'eel wall in a nanowire
with the PMA,
%where $\vec
%H_R$ is along the magnetic hard axis,
the RSOC affects the effective energy barrier in a way similar to
the adiabatic STT. Thus, its contribution to the energy barrier
appears in quadratic and higher orders of $J$.

\section*{Acknowledgments}
We acknowledge fruitful communications with Kab-Jin Kim regarding
experimental situations. This work is financially supported by the
NRF (2009-0084542, 2010-0014109, 2010-0023798,  2007-0056952, 2010-0001858), KRF
(KRF-2009-013-C00019), and BK21.

%%%%%%%%%%
%\appendix
%%%%%%%%%%
%%%%%%%%%%%%%%%%%%%%%%%%%%%%%%%%%%%%%%%%%%%%%%%%%%%%%%%%%%%%%%%%
%\section{Evaluation of disorder average} \label{sec:1D hopping}
%%%%%%%%%%%%%%%%%%%%%%%%%%%%%%%%%%%%%%%%%%%%%%%%%%%%%%%%%%%%%%%%

\end{document}